\documentclass[12pt,onecolumn]{IEEEtran}
\usepackage{mathtools}
\usepackage{multicol, nicefrac, xcolor, graphicx, relsize}
\usepackage{microtype}

\usepackage[utf8]{inputenc}
\usepackage{amsmath,amssymb,amsthm}
\usepackage{color}
\usepackage{graphicx}
\usepackage{bbm}
\usepackage{verbatim}
\usepackage{enumerate}
\usepackage{tcolorbox}
\usepackage{enumitem}
\tcbuselibrary{breakable}
\usepackage{lipsum}
\usepackage{graphicx}
\usepackage{epic}\usepackage{eepic}\usepackage{epsfig}\usepackage{amsfonts}
\usepackage{latexsym}\usepackage{color}\usepackage{enumerate}\usepackage{bbm}
\allowdisplaybreaks[4]

\usepackage[colorlinks,
linkcolor=blue,
anchorcolor=blue,
citecolor=red,
]{hyperref}

\theoremstyle{plain}
\newtheorem{theorem}{Theorem}

\newtheorem{corollary}[theorem]{Corollary}
\newtheorem{lemma}[theorem]{Lemma}
\newtheorem{proposition}[theorem]{Proposition}
\theoremstyle{definition}

\theoremstyle{remark}
\newtheorem{remark}{Remark}

\newcommand{\supp}{\operatorname{supp}}

\makeatother

\begin{document}
\title{Two-Parameter R{\'e}nyi Information Quantities with Applications to
Privacy Amplification and Soft Covering}
\author{Shi-Bing~Li, Ke~Li, and Lei~Yu 
\thanks{The work of Shi-Bing Li and Ke Li was supported by the NSFC under
grant 12031004 and grant 62571166. The work of Lei Yu was supported
by the NSFC under grant 62101286 and the Fundamental Research Funds
for the Central Universities of China (Nankai University) under grant
054-63233073. \textit{(Corresponding authors: Ke Li; Lei Yu.)}

Shi-Bing Li is with the Institute for Advanced Study
in Mathematics, School of Mathematics, Harbin Institute of Technology,
Nangang District, Harbin 150001, China (e-mail: shibingli10@gmail.com).
Ke Li is with the Institute for Advanced Study in Mathematics, Harbin
Institute of Technology, Nangang District, Harbin 150001, China (e-mail:
carl.ke.lee@gmail.com). Lei Yu is with the School of Statistics and
Data Science, LPMC, KLMDASR, and LEBPS, Nankai University, Tianjin
300071, China (e-mail: leiyu@nankai.edu.cn).}}
\maketitle

\begin{abstract}
There are no  universally accepted definitions of R{\'e}nyi conditional entropy and R{\'e}nyi mutual information, although motivated by different applications, several definitions have been proposed in the literature.
In this paper, we consider a family of two-parameter R{\'e}nyi conditional entropy and a family of two-parameter R{\'e}nyi mutual information. By performing a change of variables for the parameters, the two-parameter R{\'e}nyi conditional entropy we study coincides precisely with the definition introduced by Hayashi and Tan~\href{https://ieeexplore.ieee.org/abstract/document/7774996}{[IEEE Trans. Inf. Theory, 2016]}, and it also emerges naturally as the classical specialization of the three-parameter quantum R{\'e}nyi conditional entropy recently put forward by Rubboli, Goodarzi, and Tomamichel~\href{https://arxiv.org/abs/2410.21976}{[arXiv:2410.21976 (2024)]}. 
We establish several fundamental properties of the two-parameter R{\'e}nyi conditional entropy, including  monotonicity with respect to the parameters and variational expression.
The associated two-parameter R{\'e}nyi mutual  information considered in this paper is new and it unifies three commonly used variants of R{\'e}nyi mutual information. For this quantity, we prove several important properties, including the non-negativity, additivity, data processing inequality, monotonicity with respect to the parameters, variational expression, as well as convexity and concavity.
Finally, we demonstrate that these two-parameter R{\'e}nyi information quantities can be used to characterize the strong converse exponents in privacy amplification and soft covering problems under R{\'e}nyi divergence of order $\alpha \in (0, \infty)$.
\end{abstract}

\section{Introduction}
R{\'e}nyi entropy, introduced by A. R{\'e}nyi in 1961 \cite{Renyi1961measures}, is a 
generalization of the classical Shannon entropy and has found broad applications in 
information theory, statistics, and cryptography. Motivated by the need to analyze information-theoretic tasks across both asymptotic and non-asymptotic regimes, R{\'e}nyi entropy and R{\'e}nyi divergence offer a flexible framework for studying diverse problems such as privacy amplification, data compression, and channel resolvability \cite{Salomon2002data,Hayashi2011exponential,Hayashi2016security,YuTan2018renyi,Yu2024renyi}.

In contrast to the well-established definitions of Shannon conditional entropy and mutual information, there is no single, universally accepted formulation for R{\'e}nyi conditional entropy or R{\'e}nyi mutual information. Instead, several definitions have been proposed in the literature. Notable examples of R{\'e}nyi conditional entropy include those proposed by, e.g., Arimoto \cite{Arimoto1975information}, Cachin \cite{Cachin1997entropy}, and others \cite{GPY2009some,JizbaArimitsu2004world}. Similarly, R{\'e}nyi mutual information has been developed in various forms, e.g., by Sibson \cite{Sibson1969information}, Csiszár \cite{Csiszar1995generalized}, and Arimoto \cite{Arimoto1975information}. In response to this diversity, several recent studies \cite{TMA2012conditional,FehrBerens2014conditional,Verdu2015alpha,Aishwarya2019remarks} have sought to systematically compare these definitions, offering insights into their operational significance and mathematical properties across various regimes. Recently, in \cite{HPA2025strong}, He,   Pradhan, and  Winter introduced a novel two-parameter quantity to characterize the exact strong converse exponent for soft covering under the total variation distance.

\subsection{Our Contributions}
Our contributions are as follows. 

\begin{enumerate}
\item  The work in \cite{RGT2024quantum} introduced a two-parameter R{\'e}nyi conditional entropy that unifies two commonly adopted formulations of R{\'e}nyi conditional entropy (see also \cite{HayashiTan2016equivocations} for a different parameterization form of this quantity). We conduct a more detailed analysis of its limiting behavior as the parameters approach 0 or $\infty$, and show that it further encompasses two additional existing definitions, thereby enhancing its unifying role in the R{\'e}nyi information framework. We also propose a new definition---the two-parameter R{\'e}nyi mutual information. This information quantity, parameterized by two nonnegative numbers,  is designed to generalize several existing definitions in a unified framework. Specifically, our two-parameter R{\'e}nyi mutual information includes three widely used definitions of  R{\'e}nyi mutual information  as special cases. 

\item Beyond their unifying role, we further investigate several fundamental properties of the two-parameter quantities. These include key axiomatic characteristics such as monotonicity and continuity with respect to the R{\'e}nyi parameters, additivity, data-processing inequality, variational expression, and other structural properties. We also examine various limiting cases of the two parameters and derive explicit expressions for each case. Such properties not only enhance our theoretical understanding but also facilitate their application to practical problems in information theory and security.

\item One of the key motivations for introducing these generalized quantities lies in their applicability to strong converse analysis. In particular, we show that the two-parameter R{\'e}nyi conditional entropy and the two-parameter R{\'e}nyi mutual information can be employed to characterize the strong converse exponents in two fundamental problems: privacy amplification and soft covering. These results are derived using R{\'e}nyi divergence of order $\alpha \in (0, \infty)$ as a measure of error, thereby extending previous findings and providing a more versatile analytical tool for such settings.
\end{enumerate}

\subsection{Organization}
The remainder of this paper is organized as follows. In Section~\ref{sec:notation}, we introduce the basic notations and review several existing definitions of R{\'e}nyi conditional entropy and mutual information. Section~\ref{sec:two-Q} presents the two-parameter R{\'e}nyi conditional entropy introduced in \cite{RGT2024quantum, HayashiTan2016equivocations} and our proposed two-parameter R{\'e}nyi mutual information, establishes their relationships with existing formulations, and provides a detailed analysis of their mathematical properties. In Sections~\ref{sec:SC-PA} and~\ref{sec:SC-SC}, we apply the two-parameter information quantities to characterize the strong converse exponents in the problems of privacy amplification and soft covering, respectively. Finally, Section~\ref{sec:con-dis} concludes the paper and outlines possible directions for future research.

\begin{table*}[tbp]
% Prevent stretching above displays inside summary box
\setlength{\abovedisplayskip}{1.5ex plus0pt minus1pt}
\setlength{\belowdisplayskip}{\abovedisplayskip}
% Set-up box around summary, height is calibrated manually
\fbox{%
\begin{minipage}[t][68\baselineskip][t]{\textwidth}
\centerline{{\large\textbf{Summary}}}
\vspace{1.5\baselineskip}

\begin{multicols}{2}
\textit{Two-parameter R{\'e}nyi conditional entropy of probability distribution $P_{XY}$ for orders $\alpha\in(0,1)\cup(1,\infty)$ and $\beta\in(0,\infty)$}:
\begin{equation*}
\widetilde{H}_{\alpha,\beta}(X|Y):=  \frac{\alpha}{\beta(1-\alpha)}\log\sum_{y}P_{Y}(y)\Big(\sum_{x}P_{X|Y}^{\alpha}(x|y)\Big)^{\frac{\beta}{\alpha}}.
\end{equation*}
The cases $\beta=0,\infty$ are defined by taking the limit.

\textit{For the extended orders (Proposition~\ref{thm:conditional-entropy-definition}):}
\begin{align*}
\widetilde{H}_{0,\beta}(X|Y) &\!=\!\!
\begin{cases}
\max\limits_{y:P_{Y}(y)>0}\log \left|supp(P_{X|y})\right|, &\beta\neq0\\
\sum\limits_y P_{Y}(y)\log\left|supp(P_{X|y})\right|, &\beta=0. \\
\end{cases} \\
\widetilde{H}_{\infty,\beta}(X|Y) &\!=\!\!\begin{cases}
-\sum_{y}P_{Y}(y)\log\max\limits_{x}P_{X|Y}(x|y), & \!\!\!\beta=0\\
-\frac{1}{\beta}\!\log\sum\limits_{y}P_{Y}(y)\max\limits_{x}P_{X|Y}^{\beta}(x|y), & \!\!\!\beta\!\in\!(0,\infty) \\
-\log\max\limits_{(x,y):P_{XY}(x,y)>0}P_{X|Y}(x|y), &\!\!\! \beta=\infty.
\end{cases}
\end{align*}
\textit{Relation to existing definitions of R{\'e}nyi conditional entropy (Proposition~\ref{prop:compare-H}):} For $\alpha\in(0,1)\cup(1,\infty)$, we have
\begin{align*}
\widetilde{H}_{\alpha,\alpha}(X|Y) & =H_{\alpha}(X|Y),\\
\widetilde{H}_{\alpha,0}(X|Y) & =\bar{H}_{\alpha}(X|Y),\\
\widetilde{H}_{\alpha,1}(X|Y) & =H_{\alpha}^{*}(X|Y),\\
\widetilde{H}_{\alpha,\infty}(X|Y) & =\bar{H}_{\alpha}^{*}(X|Y).
\end{align*}

\textit{Behavior with respect to the order parameter $\alpha$ (Propositions~\ref{thm:conditional-entropy-definition}, \ref{prop:mono-alpha}, Corollary~\ref{coro:H}):}
\begin{itemize}
\item For any $\beta\in[0,\infty)$, we have
\begin{equation*}
\lim_{\alpha\to 1}\widetilde{H}_{\alpha,\beta}(X|Y)=H(X|Y).
\end{equation*}
\item For any $\beta\geq0$, $\widetilde{H}_{\alpha,\beta}(X|Y)$ is non-increasing
in $\alpha\in(0,\infty)$.
\item For any $\beta\in(0,\infty)$, $(\alpha-1)\widetilde{H}_{\alpha,\beta}(X|Y)$ is concave in $\alpha\in(0,\infty)$.
\end{itemize}
\textit{Monotonicity in $\beta$ (Proposition~\ref{prop:mono-beta}):}
\begin{itemize}
\item  When $\alpha\!\in\!(1,\infty]$, $\widetilde{H}_{\alpha,\beta}(X|Y)$
is non-increasing in $\beta\!\in\!(0,\infty)$.
\item When $\alpha\in[0,1)$, $\widetilde{H}_{\alpha,\beta}(X|Y)$
is non-decreasing in $\beta\in(0,\infty)$.
\end{itemize}
\textit{Non-negativity and Additivity (Propositions~\ref{prop:additivity}):}
\begin{itemize}
\item For any $\alpha,\beta\geq 0$, $\widetilde{H}_{\alpha,\beta}(X|Y)$ is non-negative.
\item For any $\alpha,\beta\in(0,\infty)$,
$\widetilde{H}_{\alpha,\beta}(X|Y)$ is additive.
\end{itemize}
\textit{Data processing inequality (Proposition~\ref{prop:additivity}):} Let $P_{XYZ}\in\mathcal{P}(\mathcal{X}\times \mathcal{Y}\times\mathcal{Z})$.
For any $\alpha,\beta\in(0,1]$ or $\alpha,\beta\in[1,\infty)$, we have
\begin{align*}
\widetilde{H}_{\alpha,\beta}(X|YZ) \leq\widetilde{H}_{\alpha,\beta}(X|Y).
\end{align*}
\textit{Monotonicity under discarding information (Proposition~\ref{prop:entropy-mono}):} For any $\alpha,\beta\geq0$, we
have 
\begin{equation*}
\widetilde{H}_{\alpha,\beta}(XY|Z)\geq\widetilde{H}_{\alpha,\beta}(Y|Z).
\end{equation*}
\textit{Variational expression (Theorem~\ref{thm:variational-expression-H}):} For any $\alpha,\beta\in(0,\infty)$, it holds that 
\begin{align*}
(\alpha-1)\widetilde{H}_{\alpha,\beta}(X|Y)&=\min_{Q_{XY}} \Big(\frac{\alpha(1-\beta)}{\beta}D(Q_{Y}\|P_{Y})\Big. \\
&\Big.+\alpha D(Q_{XY}\|P_{XY})+(\alpha-1)H(X|Y)\Big).
\end{align*}
\textit{Operational significance (Theorem~\ref{thm:strong-converse-exponent}):} The two-parameter R{\'e}nyi conditional entropy characterizes the strong converse exponent of privacy amplification.
\columnbreak

\textit{Two-parameter R{\'e}nyi mutual information of probability distribution $P_{XY}$ for orders $\alpha\in(0,1)\cup(1,\infty)$ and $\beta\in(0,\infty)$}:
$$\widetilde{I}_{\alpha,\beta}(X\!:\!Y)\!:= \frac{\alpha}{\beta(\alpha\!-\!1)}\log\sum_{y}P_{Y}(y)\Big(\!\sum_{x}P_{X}^{1-\alpha}(x)P_{X|Y}^{\alpha}(x|y)\Big)^{\frac{\beta}{\alpha}}\!.$$
The cases $\beta=0,\infty$ are defined by taking the limit.

\textit{For the extended orders  (Proposition~\ref{thm:mutual-definition}):} 
\begin{align*}
\widetilde{I}_{0,\beta}(X:Y)&\!=\!\!
\begin{cases}
-\max\limits_{y:P_Y(y)>0}\log\sum\limits_{x:P_{Y|X}(y|x)>0}P_{X}(x), & \beta\neq 0 \\
-\sum\limits_y P_Y(y)\log\sum\limits_{x:P_{Y|X}(y|x)>0}P_{X}(x), & \beta=0.
\end{cases} \\
\widetilde{I}_{\infty,\beta}(X\!:\!Y)&\!=\!\!\begin{cases}
\sum\limits_{y}P_{Y}(y)\log\max\limits_{x:P_{X}(x)>0}\frac{P_{X|Y}(x|y)}{P_{X}(x)}, &\!\!\!\beta=0\\
\frac{1}{\beta}\log\sum\limits_{y}P_{Y}(y)\max\limits_{x:P_{X}(x)>0}\Big(\!\frac{P_{X|Y}(x|y)}{P_{X}(x)}\!\Big)^{\beta}\!, &\!\!\!\beta\in(0,\infty)\\
\log\max\limits_{(x,y):P_{XY}(x,y)>0}\frac{P_{X|Y}(x|y)}{P_{X}(x)}, &\!\!\!\beta=\infty.
\end{cases}
\end{align*}
\textit{Relation to existing definitions of R{\'e}nyi mutual information (Proposition~\ref{prop:compare-I}):} For $\alpha\in(0,1)\cup(1,\infty)$, we have
\begin{align*}
\widetilde{I}_{\alpha,\alpha}(X:Y) & =I_{\alpha}(X:Y), \\
\widetilde{I}_{\alpha,0}(X:Y) & =\bar{I}_{\alpha}(X:Y), \\
\widetilde{I}_{\alpha,1}(X:Y) & =I_{\alpha}^{*}(X:Y),\\
\widetilde{I}_{\alpha,\infty}(X:Y) & =\bar{I}_{\alpha}^{*}(X:Y).
\end{align*}
\textit{Behavior with respect to the order parameter $\alpha$ (Propositions~\ref{thm:mutual-definition}, \ref{prop:mono-alpha-I}, Corollary~\ref{coro:I}):}
\begin{itemize}
\item For any $\beta\in[0,\infty)$, we have
\begin{equation*}
\lim_{\alpha\to1}\widetilde{I}_{\alpha,\beta}(X:Y)=I(X:Y).
\end{equation*}
\item For any $\beta\geq 0$, $\widetilde{I}_{\alpha,\beta}(X|Y)$ is non-decreasing in $\alpha \in (0,\infty)$.
\item For any $\beta\in(0,\infty)$, $(1-\alpha)\widetilde{I}_{\alpha,\beta}(X:Y)$ is concave in $\alpha\in(0,\infty)$.
\end{itemize}
\textit{Monotonicity in $\beta$ (Proposition~\ref{prop:mono-beta-I}):}
\begin{itemize}
\item  When $\alpha\in(1,\infty]$, $\widetilde{I}_{\alpha,\beta}(X:Y)$
is non-decreasing in $\beta\in(0,\infty)$.
\item  When $\alpha\in[0,1)$, $\widetilde{I}_{\alpha,\beta}(X:Y)$
is non-increasing in $\beta\in(0,\infty)$.
\end{itemize}
\textit{Non-negativity and Additivity (Propositions~\ref{prop:non-negativity}, \ref{prop:additivity-I}):}
\begin{itemize}
\item For any $\alpha,\beta\geq 0$, $\widetilde{I}_{\alpha,\beta}(X:Y)$ is non-negative.
\item For any $\alpha,\beta\in(0,\infty)$,
$\widetilde{I}_{\alpha,\beta}(X:Y)$ is additive.
\end{itemize}
\textit{Data processing inequality (Proposition~\ref{prop:DPI-I}):}
If $X-Y-Z$ is a Markov chain, we have
\begin{align*}
\widetilde{I}_{\alpha,\beta}(X:Y) & \geq\widetilde{I}_{\alpha,\beta}(X:Z), \,\alpha,\beta\in[1,\infty) \text{ or } \alpha,\beta\in(0,1].
\end{align*}
\textit{Concavity in the input distribution $P_X$ and convexity in the channel $P_{Y|X}$ (Proposition~\ref{prop:convexity-concavity}):}
\begin{itemize}
\item For fixed $P_{Y|X}$, $\widetilde{I}_{\alpha,\beta}(X:Y)$ is concave in $P_X$ for $\alpha \in[1,\infty)$ and $\beta\in(0,1]$.
\item For fixed $P_X$, $\widetilde{I}_{\alpha,\beta}(X:Y)$ is convex in $P_{Y|X}$ for $\alpha,\beta \in(0,1]$.
\end{itemize}

\textit{Variational expression (Theorem~\ref{thm:variational-expression-I}):}
For any $\alpha,\beta\in(0,\infty)$, we have 
\begin{align*}
(1\!-\alpha)\widetilde{I}_{\alpha,\beta}(X\!:\!Y)&=\!\min_{Q_{XY}}\Big(\!\frac{\alpha(1\!-\!\beta)}{\beta}D(Q_{Y}\|P_{Y})\\
&+D(Q_{XY}\|P_{XY})\!+\!(1\!-\alpha)D(Q_{X|Y}\|P_{X}|Q_{Y})\!\Big).
\end{align*}
\textit{Operational significance (Theorem~\ref{thm:SC-strong-converse-exponent}):} The two-parameter R{\'e}nyi mutual information characterizes the strong converse exponent of soft covering.
\end{multicols}
\end{minipage}
}
\end{table*}
\newpage
\section{Notation and Preliminaries}\label{sec:notation}

\subsection{Basic Notation}

Let $P_{X}$ be the probability distribution of a random variable
$X$ on alphabet $\mathcal{X}$. All alphabets considered in the sequel
are finite.
We use $\mathcal{P}(\mathcal{X})$ to denote the set of all probability
distributions on $\mathcal{X}$ and use $\supp(P_{X}):=\{x\in\mathcal{X}:P_{X}(x)\neq0\}$
to denote the support of $P_{X}\in\mathcal{P}(\mathcal{X})$. The set of conditional probability distributions on $\mathcal{Y}$ given a variable in $\mathcal{X}$ is denoted as $\mathcal{P(Y|X)}:=\{P_{Y|X}:P_{Y|X}(\cdot|x)\in\mathcal{P(Y)},\forall x\in\mathcal{X}\}$. Let
$P_{X|y}$ denote the probability distribution of $X$ given that
$Y=y$. Given $P_{X}$ and $P_{Y|X}$ , we write $P_{XY}=P_{X}P_{Y|X}$
as the joint distribution, and $P_{Y}$ as the marginal distribution of $Y$,
i.e., $P_{Y}(y)=\sum_{x}P_{X}(x)P_{Y|X}(y|x)$. For any vector $V_{X}$ on $\mathcal{X}$, define the $p$-norm
for $p\in[1,\infty)$ and $p$-quasinorm for $p\in(0,1)$, of $V_{X}$
as $\|V_{X}\|_{p}:=(\sum_{x}|V_{X}(x)|^{p})^{\frac{1}{p}}$. The $\infty$-norm of $V_X$ is defined as $\|V_{X}\|_{\infty}:=\max_x{|V_{X}(x)|}$.

We write $f(n)\dot{\leq}g(n)$ if $\limsup_{n\rightarrow\infty}\frac{1}{n}\log\frac{f(n)}{g(n)}\leq0$,
and $f(n)\dot{=}g(n)$ if both $f(n)\dot{\leq}g(n)$ and $g(n)\dot{\leq}f(n)$.
Denote $|x|^{+}:=\max\{x,0\}$ and $[n]:=\{1,2,\cdots,n\}$. Throughout this paper, the
functions $\log$ and $\exp$ are with base $2$, and $\ln$ is with
base $e$.

\subsection{R{\'e}nyi Divergence and Information Measures}
Let $P,Q\in\mathcal{P}(\mathcal{X})$. For
$\alpha\in(0,1)\cup(1,\infty)$ the order-$\alpha$ fidelity between $P$ and $Q$ is given by
\begin{equation}\label{eq:alpha-fidelity}
F_{\alpha}(P,Q):=\biggl(\sum_{x\in\mathcal{X}} P(x)^{\alpha}Q(x)^{1-\alpha}\biggr)^{\!1/(1-\alpha)}.
\end{equation}
To ensure well-definedness, when $\alpha>1$ we adopt the conventions $P^\alpha Q^{1-\alpha}=P \cdot (\tfrac{P}{Q})^{\alpha-1}$, and $\tfrac{0}{0}=0$, $\tfrac{a}{0}=\infty$ for any $a>0$. 
With the fidelity above, the order-$\alpha$ R{\'e}nyi divergence for $\alpha\in(0,1)\cup(1,\infty)$ is defined as
\begin{equation}\label{eq:Renyi-div}
D_{\alpha}(P\|Q):= -\log F_{\alpha}(P,Q).
\end{equation}
The order-$1$ R{\'e}nyi divergence is defined by taking the limit, which is equal to the relative entropy
\begin{equation}\label{eq:KL}
D(P\|Q):=\sum_{x\in\mathcal{X}} P(x)\log\frac{P(x)}{Q(x)}.
\end{equation}
Let $P_{Y|X}, Q_{Y|X}\in\mathcal{P(Y|X)}$ and $P_X\in\mathcal{P(X)}$. The conditional R{\'e}nyi divergence is defined as
\begin{equation}\label{eq:conditional-Renyi}
D_{\alpha}(P_{Y|X}\big\|Q_{Y|X}|P_X)
:=D_{\alpha}(P_XP_{Y|X}\|P_XQ_{Y|X}).
\end{equation}

For a joint probability distribution $P_{XY}\in\mathcal{P}(\mathcal{X}\times\mathcal{Y})$,
the R{\'e}nyi entropy is defined as 
\begin{align}
H_{\alpha}(X)_{P_{X}} & :=-D_{\alpha}(P_{X}\|\mathbbm{1}_{\mathcal{X}}),\label{eq:renyi-entropy}
\end{align}
where $\mathbbm{1}_{\mathcal{X}}$ is the indicator function of $\mathcal{X}$.
When $\alpha=1$, $H_{1}(X)_{P_{X}}$ is equal to the Shannon entropy
\begin{align}
H(X)_{P_{X}}:=-\sum_{x\in\mathcal{X}}P_{X}(x)\log P_{X}(x).
\end{align}
There have been several versions of the R{\'e}nyi conditional entropy
in the literature. Two primary versions are defined as 
\begin{align}
H_{\alpha}(X|Y)_{P_{XY}} & :=-D_{\alpha}(P_{XY}\|\mathbbm{1}_{\mathcal{X}}\times P_{Y}),\label{eq:renyi-condition-entropy}\\
H_{\alpha}^{*}(X|Y)_{P_{XY}} & :=-\min_{Q_{Y}\in\mathcal{P(Y)}}D_{\alpha}(P_{XY}\|\mathbbm{1}_{\mathcal{X}}\times Q_{Y}).
\end{align}
The second definition is known as Arimoto's R{\'e}nyi conditional entropy~\cite{Arimoto1975information}.
Another natural definition was introduced by Cachin~\cite{Cachin1997entropy} and later studied in~\cite{TMA2012conditional}. It is given by
\begin{align}
\bar{H}_{\alpha}(X|Y)_{P_{XY}}
:= \sum_{y\in\mathcal{Y}} P_{Y}(y) H_{\alpha}(X)_{P_{X|y}}.
\end{align}
The fourth variant discussed in~\cite{RennerWolf2005simple} is defined as
\begin{align}
\bar{H}_{\alpha}^{*}(X|Y)_{P_{XY}}
:=
\begin{cases}
\max\limits_{y: P_Y(y)>0} H_\alpha(X)_{P_{X|y}}, & \alpha \in (0,1)\\
\sum\limits_{y \in \mathcal{Y}} P_Y(y) H(X)_{P_{X|y}}, & \alpha = 1\\
\min\limits_{y: P_Y(y)>0} H_\alpha(X)_{P_{X|y}}, & \alpha \in (1,\infty).
\end{cases}
\end{align}
When $\alpha = 1$, the first three definitions of R{\'e}nyi conditional entropy are equal to the conditional entropy
\begin{align}
H(X|Y)_{P_{XY}} := \sum_{y \in \mathcal{Y}} P_Y(y) H(X)_{P_{X|y}}.
\end{align}

There have also been several versions of   R{\'e}nyi mutual information in the literature, including the following three better known versions: 
\begin{align}
I_{\alpha}(X:Y)_{P_{XY}}&:=D_{\alpha}(P_{XY}\|P_{X}\times P_{Y}),\\
I_{\alpha}^{*}(X:Y)_{P_{XY}}&:=\min_{Q_{Y}\in\mathcal{P(Y)}}D_{\alpha} (P_{XY}\|P_{X}\times Q_{Y}),\\
\bar{I}_{\alpha}(X:Y)_{P_{XY}}&:=\sum_{y\in\mathcal{Y}}P_{Y}(y)D_{\alpha}(P_{X|y}\|P_{X}).\label{eq:definition-I-1}
\end{align}
The second and third definitions are known as Sibson's R{\'e}nyi mutual information~\cite{Sibson1969information} and the Augustin-Csisz{\'a}r R{\'e}nyi mutual information~\cite{Csiszar1995generalized,Augustin1978noisy}, respectively. 
When $\alpha=1$, all these   three definitions of R{\'e}nyi mutual information 
reduce to the Shannon mutual information
\begin{equation}
I(X:Y)_{P_{XY}}:=\sum_{x\in\mathcal{X},y\in\mathcal{Y}}P_{XY}(x,y)\log \frac{P_{Y|X}(y|x)}{P_Y(y)}.
\end{equation}

\section{Two-Parameter R{\'e}nyi Information Quantities}\label{sec:two-Q}
In this section, we recall the two-parameter R{\'e}nyi conditional entropy introduced in \cite{RGT2024quantum} and \cite{HayashiTan2016equivocations}, and define a new two-parameter R{\'e}nyi mutual information. Some fundamental properties of these two quantities are established. As we will show in Sections~\ref{sec:SC-PA} and~\ref{sec:SC-SC}, these information quantities admit important applications. Specifically they play a key role in characterizing the strong converse exponents for privacy amplification and soft covering.

\subsection{Two-parameter R{\'e}nyi Conditional Entropy}
For any $\alpha\in(0,1)\cup(1,\infty)$ and $\beta\in(0,\infty)$,
a two-parameter R{\'e}nyi conditional entropy of a probability distribution
$P_{XY}\in\mathcal{P}(\mathcal{X}\times\mathcal{Y})$ is defined as \cite{RGT2024quantum, HayashiTan2016equivocations}
\begin{align}
\widetilde{H}_{\alpha,\beta}(X|Y)_{P_{XY}}:= & \frac{\alpha}{\beta(1-\alpha)}\log\sum_{y\in\mathcal{Y}}P_{Y}(y)\Big(\sum_{x\in\mathcal{X}}P_{X|Y}^{\alpha}(x|y)
\Big)^{\frac{\beta}{\alpha}}.
\end{align}
By taking   limits, we extend the definition to include the cases $\beta=0,\infty$. The following proposition shows that the two-parameter R{\'e}nyi conditional entropy encompasses other  four   existing R{\'e}nyi conditional entropies.
\begin{proposition}\label{prop:compare-H}
Let $P_{XY}\in\mathcal{P}(\mathcal{X}\times\mathcal{Y})$. For $\alpha\in(0,1)\cup(1,\infty)$, we have
\begin{align}
\widetilde{H}_{\alpha,\alpha}(X|Y)_{P_{XY}} & =H_{\alpha}(X|Y)_{P_{XY}},\label{1}\\
\widetilde{H}_{\alpha,0}(X|Y)_{P_{XY}} & =\bar{H}_{\alpha}(X|Y)_{P_{XY}},\label{2}\\
\widetilde{H}_{\alpha,1}(X|Y)_{P_{XY}} & =H_{\alpha}^{*}(X|Y)_{P_{XY}},\label{3}\\
\widetilde{H}_{\alpha,\infty}(X|Y)_{P_{XY}} & =\bar{H}_{\alpha}^{*}(X|Y)_{P_{XY}}.\label{4}
\end{align} 
\end{proposition}
Equations~\eqref{1} and \eqref{3} can be seen directly from definitions, 
whereas Equations~\eqref{2} and \eqref{4} follow from a calculation using L'H{\^o}pital's rule. In the following Proposition~\ref{thm:conditional-entropy-definition}, we fix $\beta$ and further extend the definition to the limiting cases $\alpha=0,1,\infty$.
\begin{proposition}\label{thm:conditional-entropy-definition} 
Let $P_{XY}\in\mathcal{P}(\mathcal{X}\times\mathcal{Y})$.
The following statements hold.
\begin{enumerate}
\item It holds that
\begin{equation}\label{eq:beta=00003D1}
\widetilde{H}_{1,\beta}(X|Y)_{P_{XY}}:=\lim_{\alpha\to 1}\widetilde{H}_{\alpha,\beta}(X|Y)_{P_{XY}}=H(X|Y)_{P_{XY}}, \, \beta\neq\infty.
\end{equation}
\item It holds that
\begin{align}
\widetilde{H}_{0,\beta}(X|Y)_{P_{XY}}:=\lim_{\alpha\to 0}\widetilde{H}_{\alpha,\beta}(X|Y)_{P_{XY}} =\left\{
\begin{array}{ll}
\max\limits_{y:P_{Y}(y)>0}\log \left|\supp(P_{X|y})\right|, &\beta\neq0\\
\sum\limits_y P_{Y}(y)\log\left|\supp(P_{X|y})\right|, &\beta=0.\label{eq:beta=00003D0}
\end{array}
\right.
\end{align}
\item We have
\begin{align}
\widetilde{H}_{\infty,\beta}(X|Y)_{P_{XY}} :=\lim_{\alpha\to \infty}\widetilde{H}_{\alpha,\beta}(X|Y)_{P_{XY}}=\begin{cases}
-\sum\limits_{y}P_{Y}(y)\log\max\limits_{x}P_{X|Y}(x|y), & \beta=0\\
-\frac{1}{\beta}\log\sum\limits_{y}P_{Y}(y)\max\limits_{x}P_{X|Y}^{\beta}(x|y), & \beta\in(0,\infty) \\
-\log\max\limits_{(x,y):P_{XY}(x,y)>0}P_{X|Y}(x|y), & \beta=\infty.
\end{cases}
\end{align}
\end{enumerate}
\end{proposition}
\begin{IEEEproof}
For any $\beta\neq\infty$, it is straightforward to verify by L'H{\^o}pital's rule that 
\begin{equation}
\lim_{\alpha\to 1}\widetilde{H}_{\alpha,\beta}(X|Y)_{P_{XY}}=H(X|Y)_{P_{XY}}.
\end{equation}
Next, we prove Statement 2. When $\beta=0$ and $\infty$, the desired results follow directly from the definition. Suppose that $\beta\in(0,\infty)$.
For any $\epsilon>0$, there exists a sufficiently small $\alpha>0$
such that 
\begin{equation}
\left|\supp(P_{X|y})\right|\cdot\boldsymbol{1}{\{P_{Y}(y)>0\}}-\epsilon<\sum_{x}P_{X|Y}^{\alpha}(x|y)P_{Y}^{\frac{\alpha}{\beta}}(y)\leq\left|\supp(P_{X|y})\right|\cdot\boldsymbol{1} {\{P_{Y}(y)>0\}}.
\end{equation}
Since $\epsilon>0$ is arbitrary, using the ${\infty}$-norm gives
\begin{align}
& \lim_{\alpha\to0}\frac{1}{1-\alpha}\log\Big(\sum_{y}\Big(\sum_{x}P_{X|Y}^{\alpha}(x|y)P_{Y}^{\frac{\alpha}{\beta}}(y)\Big)^{\frac{\beta}{\alpha}}\Big)^{\frac{\alpha}{\beta}}\nonumber \\
= & \max_{y:P_{Y}(y)>0}\log\left|\supp(P_{X|y})\right|.
\end{align}
This completes the proof of Statement 2. 
Statement 3 can be obtained directly by a simple calculation.
\end{IEEEproof}

\begin{remark} In fact, the definition of $\widetilde{H}_{\alpha,\beta}(X|Y)_{P_{XY}}$
here is not continuous at $(\alpha,\beta)=(0,0)$. For example, 
\begin{align}
\lim_{\alpha=\beta\to0}\widetilde{H}_{\alpha,\beta}(X|Y)_{P_{XY}}=\log\sum_{y}P_{Y}(y)\left|\supp(P_{X|y})\right|\geq\widetilde{H}_{0,0}(X|Y)_{P_{XY}},
\end{align}
where the inequality can be strict. Our definition
of $\widetilde{H}_{0,0}(X|Y)_{P_{XY}}$ is obtained by taking limit
$\beta\to0$ first and $\alpha\to0$ then. Taking limits along different
paths could yield  different variant definitions of $\widetilde{H}_{0,0}(X|Y)_{P_{XY}}$.
\end{remark}

The following proposition states that the two-parameter R{\'e}nyi conditional entropy is non-negative, additive and satisfies the data processing inequality, which were previously established in a more general framework of the quantum setting in \cite{RGT2024quantum}.
\begin{proposition}\label{prop:additivity}
The following statements holds.
\begin{enumerate}
\item Let $P_{XY}\in\mathcal{P}(\mathcal{X}\times\mathcal{Y})$. For any $\alpha,\beta\in(0,\infty)$, $\widetilde{H}_{\alpha,\beta}(X|Y)_{P_{XY}}$ is non-negative.
\item Let $P_{XY}\in\mathcal{P}(\mathcal{X}\times\mathcal{Y})$ and $Q_{X'Y'}\in\mathcal{P}(\mathcal{X'}\times\mathcal{Y'})$. For any $\alpha,\beta\in(0,\infty)$, we have
\begin{equation}
\widetilde{H}_{\alpha,\beta}(XX'|YY')_{P_{XY}\times Q_{X'Y'}}=\widetilde{H}_{\alpha,\beta}(X|Y)_{P_{XY}}+\widetilde{H}_{\alpha,\beta}(X'|Y')_{Q_{X'Y'}}.
\end{equation}
\item Let $P_{XYZ}\in\mathcal{P}(\mathcal{X}\times\mathcal{Y}\times\mathcal{Z})$. For any $\alpha,\beta\in(0,1]$ or $\alpha,\beta\in[1,\infty)$, we have
\begin{align}
\widetilde{H}_{\alpha,\beta}(X|YZ)_{P_{XYZ}} \leq&\widetilde{H}_{\alpha,\beta}(X|Y)_{P_{XY}}. \label{eq:DPI-H-1}
\end{align}
\end{enumerate} 
\end{proposition}

Propositions~\ref{prop:mono-alpha} and~\ref{prop:mono-beta} address
the monotonicity of the two-parameter R{\'e}nyi conditional entropy with respect to one parameter when the other is held fixed.
\begin{proposition}[Monotonicity in $\alpha$]\label{prop:mono-alpha}
Let $P_{XY}\in\mathcal{P}(\mathcal{X}\times\mathcal{Y})$. For any
$\beta\geq0$, $\widetilde{H}_{\alpha,\beta}(X|Y)_{P_{XY}}$ is non-increasing
in $\alpha\in(0,\infty)$.
\end{proposition}
\begin{IEEEproof}
When $\beta=0$ or $\infty$, this is proven in \cite{TMA2012conditional}. So, we only need to consider the case $\beta\in(0,\infty)$.
Let $a,b\in(1,\infty)$ with $a\geq b$. We will show that $\widetilde{H}_{a,\beta}(X|Y)_{P_{XY}}\leq\widetilde{H}_{b,\beta}(X|Y)_{P_{XY}}$. Since the two-parameter R{\'e}nyi conditional entropy can be rewritten as
\begin{equation}
\widetilde{H}_{\alpha,\beta}(X|Y)_{P_{XY}}=\frac{\alpha}{\beta(1-\alpha)}\log\sum_{y}P_{Y}(y)\|P_{X|y}\|_{\alpha}^{\beta},
\end{equation}
this is equivalent to
\begin{equation}
\Big(\sum_{y}P_{Y}(y)\|P_{X|y}\|_{a}^{\beta}\Big)^{\frac{a}{\beta(a-1)}}\geq\Big(\sum_{y}P_{Y}(y)\|P_{X|y}\|_{b}^{\beta}\Big)^{\frac{b}{\beta(b-1)}}.
\end{equation}
Obviously, $\frac{a(b-1)}{b(a-1)}\leq1$ and $\frac{b-1}{a-1}\leq1$.
Using Jensen's inequality, we have 
\begin{align}
\Big(\sum_{y}P_{Y}(y)\|P_{X|y}\|_{a}^{\beta}\Big)^{\frac{a}{\beta(a-1)}} & =\Big(\sum_{y}P_{Y}(y)\|P_{X|y}\|_{a}^{\beta}\Big)^{\frac{b}{\beta(b-1)}\cdot\frac{a(b-1)}{b(a-1)}}\nonumber \\
& \geq\Big(\sum_{y}P_{Y}(y)\|P_{X|y}\|_{a}^{\beta\cdot\frac{a(b-1)}{b(a-1)}}\Big)^{\frac{b}{\beta(b-1)}}\nonumber \\
& =\Big(\sum_{y}P_{Y}(y)\Big(\sum_{x}P_{X|Y}(x|y)P_{X|Y}^{a-1}(x|y)\Big)^{\frac{\beta}{b}\cdot\frac{b-1}{a-1}}\Big)^{\frac{b}{\beta(b-1)}}\nonumber \\
& \geq\Big(\sum_{y}P_{Y}(y)\Big(\sum_{x}P_{X|Y}(x|y)P_{X|Y}^{b-1}(x|y)\Big)^{\frac{\beta}{b}}\Big)^{\frac{b}{\beta(b-1)}}\nonumber \\
& =\Big(\sum_{y}P_{Y}(y)\|P_{X|y}\|_{b}^{\beta}\Big)^{\frac{b}{\beta(b-1)}}.
\end{align}
This leads to the monotonicity for $\alpha\in(1,\infty)$. For $\alpha\in(0,1)$,
the desired result follows by similar arguments. Finally, since the function is continuous at $\alpha=1$, we complete the proof.
\end{IEEEproof}
\begin{proposition}[Monotonicity in $\beta$]\label{prop:mono-beta}
Let $P_{XY}\in\mathcal{P}(\mathcal{X}\times\mathcal{Y})$. The following
statements hold.
\begin{enumerate}
\item  When $\alpha\in(1,\infty]$, $\widetilde{H}_{\alpha,\beta}(X|Y)_{P_{XY}}$
is non-increasing in $\beta\in(0,\infty)$.
\item When $\alpha\in[0,1)$, $\widetilde{H}_{\alpha,\beta}(X|Y)_{P_{XY}}$
is non-decreasing in $\beta\in(0,\infty)$.
\end{enumerate}
\end{proposition}
In the above proposition, we do not include the situation $\alpha=1$, because $\widetilde{H}_{1,\beta}(X|Y)_{P_{XY}}$ reduces to the conditional entropy for any $\beta\in(0,\infty)$.
\begin{IEEEproof}[Proof of Proposition~\ref{prop:mono-beta}]
Let $\infty>a\geq b>0$. When $\alpha\in(1,\infty)$, using Jensen's inequality, we have
\begin{align}
\Big(\sum_{y}P_{Y}(y)\|P_{X|y}\|_{\alpha}^{b}\Big)^{\frac{\alpha}{b(\alpha-1)}} 
& =\Big(\sum_{y}P_{Y}(y)\|P_{X|y}\|_{\alpha}^{b}\Big)^{\frac{\alpha}{a(\alpha-1)}\cdot\frac{a}{b}}\nonumber\\
& \leq\Big(\sum_{y}P_{Y}(y)\|P_{X|y}\|_{\alpha}^{a}\Big)^{\frac{\alpha}{a(\alpha-1)}}.
\end{align}
When $\alpha\in(0,1)$, we have
\begin{equation}
\Big(\sum_{y}P_{Y}(y)\|P_{X|y}\|_{\alpha}^{b}\Big)^{\frac{\alpha}{b(\alpha-1)}}  \geq\Big(\sum_{y}P_{Y}(y)\|P_{X|y}\|_{\alpha}^{a}\Big)^{\frac{\alpha}{a(\alpha-1)}}.
\end{equation}
Hence, the desired monotonicity with respect to $\beta \in (0,\infty)$ holds for each fixed $\alpha \in (0,1) \cup (1,\infty)$.
The cases $\alpha=0$ and $\alpha=\infty$ can be obtained by taking the limit.
\end{IEEEproof}

The following proposition establishes the monotonicity of the two-parameter R{\'e}nyi conditional entropy with respect to the number of random variables. It is crucial in our proof of the optimality part of the strong converse exponent for privacy amplification in Section~\ref{sec:SC-PA}.

\begin{proposition}\label{prop:entropy-mono} 
Let $P_{XYZ}\in\mathcal{P}(\mathcal{X}\times\mathcal{Y}\times\mathcal{Z})$. For any $\alpha,\beta\in(0,\infty)$, we
have 
\begin{equation}
\widetilde{H}_{\alpha,\beta}(XY|Z)_{P_{XYZ}}\geq\widetilde{H}_{\alpha,\beta}(Y|Z)_{P_{YZ}}.
\end{equation}
\end{proposition}
\begin{IEEEproof}
Suppose that $\alpha\in(0,1)\cup(1,\infty)$. We have 
\begin{align}
\widetilde{H}_{\alpha,\beta}(XY|Z)_{P_{XYZ}} & =\frac{\alpha}{\beta(1-\alpha)}\log\sum_{z}P_{Z}(z)\Big(\sum_{x,y}P_{XY|Z}^{\alpha}(x,y|z)\Big)^{\frac{\beta}{\alpha}}\nonumber \\
& \geq\frac{\alpha}{\beta(1-\alpha)}\log\sum_{z}P_{Z}(z)\Big(\sum_{y}\Big(\sum_{x}P_{XY|Z}(x,y|z)\Big)^{\alpha}\Big)^{\frac{\beta}{\alpha}}\nonumber \\
& =\frac{\alpha}{\beta(1-\alpha)}\log\sum_{z}P_{Z}(z)\Big(\sum_{y}P_{Y|Z}^{\alpha}(y|z)\Big)^{\frac{\beta}{\alpha}}\nonumber \\
& =\widetilde{H}_{\alpha,\beta}(Y|Z)_{P_{YZ}},
\end{align}
where the inequality follows from Lemma~\ref{lem:-2}. The case $\alpha = 1$ follows directly from taking the limit.
\end{IEEEproof}

In the following Theorem \ref{thm:variational-expression-H}, we derive
a variational expression for the two-parameter R{\'e}nyi conditional
entropy, relating it to the ordinary relative entropy and conditional entropy.
\begin{theorem}[Variational Expression]\label{thm:variational-expression-H} 
Let $P_{XY}\in\mathcal{P}(\mathcal{X}\times\mathcal{Y})$. For any $\alpha,\beta\in(0,\infty)$, it holds that 
\begin{equation}
(\alpha-1)\widetilde{H}_{\alpha,\beta}(X|Y)_{P_{XY}}=\min_{Q_{XY}\in\mathcal{P}(\mathcal{X}\times \mathcal{Y})}\Big\{\frac{\alpha(1-\beta)}{\beta}D(Q_{Y}\|P_{Y})+\alpha D(Q_{XY}\|P_{XY})+(\alpha-1)H(X|Y)_{Q_{XY}}\Big\}.\label{eq:variational-H}
\end{equation}
\end{theorem}
\begin{IEEEproof}
When $\alpha=1$, the left hand side of Eq.~\eqref{eq:variational-H} is equal to 0. Using the data processing inequality of relative entropy, we have
\begin{equation}
\frac{1-\beta}{\beta}D(Q_{Y}\|P_{Y})+D(Q_{XY}\|P_{XY})\geq \frac{1}{\beta}D(Q_{Y}\|P_{Y}).
\end{equation}
Thus, the right hand side of Eq.~\eqref{eq:variational-H} is non-negative and it attains zero when we set $Q_{XY}=P_{XY}$. Therefore,  both sides of Eq.~\eqref{eq:variational-H} are equal to 0.

Consider the case $\alpha\in(0,1)\cup(1,\infty)$ and $\beta\in(0,\infty)$.
The definition of the two-parameter R{\'e}nyi conditional entropy provides that
\begin{equation}
(\alpha-1)\widetilde{H}_{\alpha,\beta}(X|Y)_{P_{XY}}
= -\frac{\alpha}{\beta}\log\sum_{y}P_{Y}(y)\Big(\sum_{x}P_{X|Y}^{\alpha}(x|y)
\Big)^{\frac{\beta}{\alpha}}.\label{eq:VE-H-1}
\end{equation}
Define a probability distribution
\begin{equation}
\widetilde{Q}_Y(y)= \frac{P_Y(y)(\sum_{x}P_{X|Y}^{\alpha}(x|y))^{\frac{\beta}{\alpha}}}{\sum_{y'}P_{Y}(y')(\sum_{x'}P_{X|Y}^{\alpha}(x'|y'))^{\frac{\beta}{\alpha}}},
\end{equation} 
and for each $y\in\mathcal{Y}$ define a probability distribution
\begin{equation}
\widetilde{Q}_X^{(y)}(x)= \frac{P_{X|Y}^{\alpha}(x|y)}{\sum_{x'}P_{X|Y}^{\alpha}(x'|y)}.
\end{equation}
Then, we have
\begin{align}
& -\log\sum_{y}P_{Y}(y)\Big(\sum_{x}P_{X|Y}^{\alpha}(x|y)
\Big)^{\frac{\beta}{\alpha}}\nonumber \\
= &\min_{Q_{Y}\in\mathcal{P}( \mathcal{Y})} \Big\{D(Q_Y\|\widetilde{Q}_Y) -\log\sum_{y}P_{Y}(y)\Big(\sum_{x}P_{X|Y}^{\alpha}(x|y)
\Big)^{\frac{\beta}{\alpha}}\Big\} \nonumber\\
= &\min_{Q_{Y}\in\mathcal{P}( \mathcal{Y})}\Big\{D(Q_{Y}\|P_{Y})-\frac{\beta}{\alpha}\sum_{y}Q_Y(y)\log\sum_{x}P_{X|Y}^{\alpha}(x|y)\Big\}\nonumber \\
= &\min_{Q_{Y}\in\mathcal{P}( \mathcal{Y})}\Big\{D(Q_{Y}\|P_{Y})+\frac{\beta}{\alpha}\sum_{y}Q_Y(y)\Big[\min_{Q_{X}^{(y)}\in\mathcal{P(X)}}\Big(D(Q_{X}^{(y)}\|\widetilde{Q}_X^{(y)})-\log\sum_{x}P_{X|Y}^{\alpha}(x|y)\Big)\Big]\Big\}\nonumber \\
= &\min_{Q_{Y}\in\mathcal{P}( \mathcal{Y})}\Big\{D(Q_{Y}\|P_{Y})-\frac{\beta}{\alpha}\sum_{y}Q_Y(y)\Big[\min_{Q_{X}^{(y)}\in\mathcal{P(X)}}\Big(H(X)_{Q_{X}^{(y)}}+\alpha\sum_{x}Q_{X}^{(y)}(x)\log P_{X|Y}(x|y)\Big)\Big]\Big\} \nonumber\\
=&\min_{Q_{Y}\in\mathcal{P}( \mathcal{Y})}\Big\{D(Q_{Y}\|P_{Y})-\frac{\beta}{\alpha}\min_{Q_{X|Y}\in\mathcal{P}(\mathcal{X}|\mathcal{Y})}\Big(H(X|Y)_{Q_{XY}}+\alpha\sum_{x,y}Q_{XY}(x,y)\log P_{X|Y}(x|y)\Big)\Big\} \nonumber\\
= & \min_{Q_{XY}\in\mathcal{P}(\mathcal{X}\times \mathcal{Y})}\Big\{D(Q_{Y}\|P_{Y})-\frac{\beta}{\alpha}H(X|Y)_{Q_{X|Y}}-\beta\sum_{x,y}Q_{XY}(x,y)\log P_{X|Y}(x|y)\Big\},\label{eq:VE-H-2}
\end{align}
where in the fifth equality we have identified $Q_{X|Y}(\cdot|y)$ with $Q_X^{(y)} $. It can be verified by direct calculation that
\begin{equation}
\sum_{x,y}Q_{XY}(x,y)\log P_{X|Y}(x|y)=D(Q_{Y}\|P_{Y})-D(Q_{XY}\|P_{XY})-H(X|Y)_{Q_{XY}}.\label{eq:VE-H-3}
\end{equation}
Combining Eqs.~\eqref{eq:VE-H-1}, \eqref{eq:VE-H-2} and \eqref{eq:VE-H-3} yields the claimed result.
\end{IEEEproof}
Theorem~\ref{thm:variational-expression-H} shows that $(\alpha-1)\widetilde{H}_{\alpha,\beta}(X|Y)_{P_{XY}}$ is the minimization over a set of functions that are linear in $\alpha\in(0,\infty)$. This directly implies the following corollary.
\begin{corollary}\label{coro:H}
For any $\beta\in(0,\infty)$, the function $(\alpha-1)\widetilde{H}_{\alpha,\beta}(X|Y)_{P_{XY}}$ is concave in $\alpha$ on $(0,\infty)$.
\end{corollary}

\subsection{Two-parameter R{\'e}nyi Mutual Information}
We now introduce a new version of R{\'e}nyi mutual information, which
is coined by us  as the \emph{two-parameter R{\'e}nyi mutual information}.
Given a probability distribution $P_{XY}\in\mathcal{P}(\mathcal{X}\times\mathcal{Y})$,
for any $\alpha\in(0,1)\cup(1,\infty)$ and $\beta\in(0,\infty)$,
our two-parameter R{\'e}nyi mutual information is defined as 
\begin{align}
\widetilde{I}_{\alpha,\beta}(X:Y)_{P_{XY}}:=\frac{\alpha}{\beta(\alpha-1)}\log\sum_{y\in\mathcal{Y}}P_{Y}(y)\Big(\sum_{x\in\mathcal{X}}P_{X}^{1-\alpha}(x)P_{X|Y}^{\alpha}(x|y)\Big)^{\frac{\beta}{\alpha}}.
\end{align}
By taking limits, we extend the definition to include the cases $\beta=0,\infty$.
The following proposition shows that the two-parameter
R{\'e}nyi mutual information encompasses three existing versions of R{\'e}nyi
mutual information, as well as the following new one:  
\begin{align}
\bar{I}_{\alpha}^{*}(X:Y)_{P_{XY}}:=\begin{cases}
\min\limits _{y:P_{Y}(y)>0}D_{\alpha}(P_{X|y}\|P_{X}), & \alpha\in(0,1)\\
I(X:Y)_{P_{XY}}, & \alpha=1\\
\max\limits _{y:P_{Y}(y)>0}D_{\alpha}(P_{X|y}\|P_{X}), & \alpha\in(1,\infty).
\end{cases}
\end{align}
\begin{proposition}\label{prop:compare-I} Let $P_{XY}\in\mathcal{P}(\mathcal{X\times\mathcal{Y}})$.
For any $\alpha\in(0,1)\cup(1,\infty)$, we have 
\begin{align}
\widetilde{I}_{\alpha,\alpha}(X:Y)_{P_{XY}} & =I_{\alpha}(X:Y)_{P_{XY}},\label{11}\\
\widetilde{I}_{\alpha,0}(X:Y)_{P_{XY}} & =\bar{I}_{\alpha}(X:Y)_{P_{XY}},\label{12}\\
\widetilde{I}_{\alpha,1}(X:Y)_{P_{XY}} & =I_{\alpha}^{*}(X:Y)_{P_{XY}},\label{13}\\
\widetilde{I}_{\alpha,\infty}(X:Y)_{P_{XY}} & =\bar{I}_{\alpha}^{*}(X:Y)_{P_{XY}}.\label{14}
\end{align}
\end{proposition} Equations~\eqref{11} and \eqref{13} come directly
from definitions, whereas Equations~\eqref{12} and \eqref{14} follow
from a calculation using L'Hôpital's rule.  In Proposition~\ref{thm:mutual-definition} below we fix $\beta$ and further extend the definition to the limiting cases $\alpha=0,1,\infty$.
\begin{proposition}\label{thm:mutual-definition} 
Let $P_{XY}\in\mathcal{P}(\mathcal{X}\times\mathcal{Y})$. The following statements hold.
\begin{enumerate}
\item It holds that
\begin{equation}
\widetilde{I}_{1,\beta}(X:Y)_{P_{XY}}:=\lim_{\alpha\to1}\widetilde{I}_{\alpha,\beta}(X:Y)_{P_{XY}}=I(X:Y)_{P_{XY}},\,\beta\neq\infty.\label{eq:I-beta=00003D1}
\end{equation}
\item We have
\begin{align}
\widetilde{I}_{0,\beta}(X:Y)_{P_{XY}}:=\lim_{\alpha\to0}\widetilde{I}_{\alpha,\beta}(X:Y)_{P_{XY}}=\left\{
\begin{array}{ll}
-\max\limits_{y: P_Y(y)>0}\log\sum\limits_{x:P_{Y|X}(y|x)>0}P_{X}(x), & \beta\neq 0 \\
-\sum\limits_y P_Y(y)\log\sum\limits_{x:P_{Y|X}(y|x)>0}P_{X}(x), & \beta=0.
\end{array}
\right. 
\end{align}
\item It holds that
\begin{align}
\widetilde{I}_{\infty,\beta}(X:Y)_{P_{XY}}:=\lim_{\alpha\to\infty}\widetilde{I}_{\alpha,\beta}(X:Y)_{P_{XY}}=\left\{
\begin{array}{ll}
\sum\limits_{y}P_{Y}(y)\log\max\limits_{x:P_{X}(x)>0}\frac{P_{X|Y}(x|y)}{P_{X}(x)}, &\beta=0\\
\frac{1}{\beta}\log\sum\limits_{y}P_{Y}(y)\max\limits_{x:P_{X}(x)>0}\Big(\frac{P_{X|Y}(x|y)}{P_{X}(x)}\Big)^{\beta}, &\beta\in(0,\infty)\\
\log\max\limits_{(x,y):P_{XY}(x,y)>0}\frac{P_{X|Y}(x|y)}{P_{X}(x)}, &\beta=\infty.
\end{array}\right.
\end{align}
\end{enumerate}
\end{proposition}
\begin{IEEEproof}
Due to L'H{\^o}pital's rule, it is easy to verify that
\begin{align}
\lim_{\alpha\to1}\widetilde{I}_{\alpha,\beta}(X:Y)_{P_{XY}}=I(X:Y)_{P_{XY}}.
\end{align}
We now proceed to prove Statement 2. When $\beta=0$ and $\infty$, the desired results follow directly from the definition. Consider the case $\beta \in (0, \infty)$. When $P_Y(y)>0$, for
any $\epsilon>0$, there exists a sufficiently small $\alpha>0$ such
that 
\begin{equation}
\sum_{x:P_{Y|X}(y|x)>0}P_{X}(x)-\epsilon<\sum_{x}P_{X}(x)P_{Y|X}^{\alpha}(y|x)P_{Y}^{\frac{1-\beta}{\beta}\cdot\alpha}(y)<\sum_{x:P_{Y|X}(y|x)>0}P_{X}(x)+\epsilon.
\end{equation}
Since $\epsilon>0$ is arbitrary, using $\infty$-norm gives
\begin{align}
&\lim_{\alpha\to0}\frac{1}{\alpha-1}\log\Big(\sum_{y}\Big(\sum_{x}P_{X}(x)P_{Y|X}^{\alpha}(y|x)P_{Y}^{\frac{1-\beta}{\beta}\cdot\alpha}(y)\Big)^{\frac{\beta}{\alpha}}\Big)^{\frac{\alpha}{\beta}}\nonumber\\
=&-\max_{y:P_Y(y)>0}\log\sum_{x:P_{Y|X}(y|x)>0}P_{X}(x).
\end{align}
The proof of Statement~2 is complete.
Statement 3 follows directly from a simple calculation.
\end{IEEEproof}

\begin{remark}
Similarly to the condition entropy
case, $\widetilde{I}_{\alpha,\beta}(X|Y)_{P_{XY}}$ is not continuous
at $(\alpha,\beta)=(0,0)$, e.g.,
\begin{align}
\lim_{\alpha=\beta\to0}\widetilde{I}_{\alpha,\beta}(X|Y)_{P_{XY}}=-\log\sum\limits_{(x,y):P_{XY}(x,y)>0}P_{X}(x)P_{Y}(y)\leq\widetilde{I}_{0,0}(X|Y)_{P_{XY}},
\end{align}
where the inequality can be strict. Taking limits along different
paths could yield different  variant definitions of $\widetilde{I}_{0,0}(X|Y)_{P_{XY}}$.
\end{remark} 

In the following proposition, we establish the non-negativity of the two-parameter R{\'e}nyi mutual information.
\begin{proposition}[Non-Negativity]\label{prop:non-negativity}
Let $P_{XY}\in\mathcal{P}(\mathcal{X}\times\mathcal{Y})$. For any $\alpha,\beta\geq0$, $\widetilde{I}_{\alpha,\beta}(X:Y)_{P_{XY}}\geq0$, with equality if and only if $X$ and $Y$ are independent.
\end{proposition}
\begin{IEEEproof}
It is enough to consider $\alpha\in(0,1)\cup(1,\infty)$ and $\beta\in(0,\infty)$, as the other cases follow easily by using corresponding limits. When $\alpha\in(0,1)$, using Jensen's inequality, we have
\begin{align}
\sum_{x}P_{X}^{1-\alpha}(x)P_{X|Y}^{\alpha}(x|y)
=&\sum_x P_X(x)\Big(\frac{P_{X|Y}(x|y)}{P_X(x)}\Big)^\alpha \nonumber\\
\leq& \Big(\sum_x P_{X|Y}(x|y)\Big)^\alpha=1.
\end{align}
When $\alpha\in(1,\infty)$, we have
\begin{equation}
\sum_{x}P_{X}^{1-\alpha}(x)P_{X|Y}^{\alpha}(x|y)
\geq \Big(\sum_x P_{X|Y}(x|y)\Big)^\alpha=1.
\end{equation}
This directly implies the non-negativity. From the equality condition of Jensen's inequality, we obtain $\widetilde{I}_{\alpha,\beta}(X:Y)_{P_{XY}}=0$ if and only if $X$ and $Y$ are independent.
\end{IEEEproof}
In Proposition~\ref{prop:additivity-I}, we prove the additivity of the two-parameter R{\'e}nyi mutual information. Propositions~\ref{prop:mono-alpha-I} and~\ref{prop:mono-beta-I} establish its monotonicity with respect to each parameter individually, holding the other fixed.
\begin{proposition}[Additivity]\label{prop:additivity-I}
Let $P_{XY}\in\mathcal{P}(\mathcal{X}\times\mathcal{Y})$ and $Q_{X'Y'}\in\mathcal{P}(\mathcal{X'}\times\mathcal{Y'})$.
For $\alpha,\beta\in(0,\infty)$, we have
\begin{equation}
\widetilde{I}_{\alpha,\beta}(XX':YY')_{P_{XY}\times Q_{X'Y'}}=\widetilde{I}_{\alpha,\beta}(X:Y)_{P_{XY}}+\widetilde{I}_{\alpha,\beta}(X':Y')_{Q_{X'Y'}}.
\end{equation}
\end{proposition}
\begin{IEEEproof}
We can write
\begin{equation}
\widetilde{I}_{\alpha,\beta}(X:Y)_{P_{XY}}= \frac{\alpha}{\beta(\alpha-1)}\log\sum_y P_Y(y)\|V_{X|y}\|_\alpha^\beta,
\end{equation}
where
\begin{equation}
V_{X|y}(x)=P_X^{\frac{1-\alpha}{\alpha}}(x)P_{X|Y}(x|y).\label{eq:vector}
\end{equation}
It is easy to check that 
\begin{equation}
\sum_{y,y'} P_Y(y)Q_{Y'}(y')\|V_{X|y}\times U_{X'|y'}\|_\alpha^\beta=\Big(\sum_{y} P_Y(y)\|V_{X|y}\|_\alpha^\beta\Big)\cdot \Big(\sum_{y'} Q_{Y'}(y')\| U_{X'|y'}\|_\alpha^\beta\Big),
\end{equation}
where 
\begin{equation}
U_{X'|y'}(x')=P_{X'}^{\frac{1-\alpha}{\alpha}}(x')P_{X'|Y'}(x'|y').
\end{equation}
So, the additivity follows.
\end{IEEEproof}
\begin{proposition}[Monotonicity in $\alpha$]\label{prop:mono-alpha-I}
Let $P_{XY}\in\mathcal{P}(\mathcal{X}\times\mathcal{Y})$.
For any $\beta\geq 0$, $\widetilde{I}_{\alpha,\beta}(X:Y)_{P_{XY}}$ is non-decreasing in $\alpha \in (0,\infty)$.
\end{proposition}
\begin{IEEEproof}
The proof is similar to that of Proposition~\ref{prop:mono-alpha}. It suffices to prove the case $\beta\in(0,\infty)$, since the cases $\beta=0$ and $\infty$ can be handled by taking the limit. Let $a,b\in(1,\infty)$ with $a\geq b$. We will show that
\begin{equation}
\widetilde{I}_{a,\beta}(X:Y)_{P_{XY}}\geq\widetilde{I}_{b,\beta}(X:Y)_{P_{XY}}.
\end{equation} 
Equivalently,
\begin{equation}
\Big(\sum_{y}P_{Y}(y)\|V_{X|y}\|_{a}^{\beta}\Big)^{\frac{a}{\beta(a-1)}}\geq\Big(\sum_{y}P_{Y}(y)\|V_{X|y}\|_{b}^{\beta}\Big)^{\frac{b}{\beta(b-1)}},
\end{equation}
where $V_{X|y}$ is defined in Eq.~\eqref{eq:vector}.
Since $a\geq b>1$, we have $\frac{a(b-1)}{b(a-1)}\leq1$ and $\frac{b-1}{a-1}\leq1$.
Using Jensen's inequality, we get 
\begin{align}
&\Big(\sum_{y}P_{Y}(y)\|V_{X|y}\|_{a}^{\beta}\Big)^{\frac{a}{\beta(a-1)}} \nonumber\\
=&\Big(\sum_{y}P_{Y}(y)\Big(\sum_{x}P_{X}^{1-a}(x)P_{X|Y}^{a}(x|y)\Big)^{\frac{\beta}{a}}\Big)^{\frac{b}{\beta(b-1)}\cdot\frac{a(b-1)}{b(a-1)}}\nonumber\\
\geq&\Big(\sum_{y}P_{Y}(y)\Big(\sum_{x}P_{X}^{1-a}(x)P_{X|Y}^{a}(x|y)\Big)^{\frac{b-1}{a-1}\cdot\frac{\beta}{b}}\Big)^{\frac{b}{\beta(b-1)}}\nonumber\\
=&\Big(\sum_{y}P_{Y}(y)\Big(\sum_{x}P_{X|Y}(x|y)P_{X}^{1-a}(x)P_{X|Y}^{a-1}(x|y)\Big)^{\frac{b-1}{a-1}\cdot\frac{\beta}{b}}\Big)^{\frac{b}{\beta(b-1)}}\nonumber\\
\geq&\Big(\sum_{y}P_{Y}(y)\Big(\sum_{x}P_{X|Y}(x|y)P_{X}^{1-b}(x)P_{X|Y}^{b-1}(x|y)\Big)^{\frac{\beta}{b}}\Big)^{\frac{b}{\beta(b-1)}}\nonumber\\
=&\Big(\sum_{y}P_{Y}(y)\Big(\sum_{x}P_{X}^{1-b}(x)P_{X|Y}^{b}(x|y)\Big)^{\frac{\beta}{b}}\Big)^{\frac{b}{\beta(b-1)}}.
\end{align}
This establishes the monotonicity in $\alpha\in(1,\infty)$. For $\alpha\in(0,1)$ the desired result follows by similar arguments. Finally, since the function is continuous at $\alpha=1$, the proof is complete.
\end{IEEEproof}
\begin{proposition}[Monotonicity in $\beta$]\label{prop:mono-beta-I}
Let $P_{XY}\in\mathcal{P}(\mathcal{X}\times\mathcal{Y})$. The following
statements hold.
\begin{enumerate}
\item  When $\alpha\in(1,\infty]$, $\widetilde{I}_{\alpha,\beta}(X:Y)_{P_{XY}}$
is non-decreasing in $\beta\in(0,\infty)$.
\item  When $\alpha\in[0,1)$, $\widetilde{I}_{\alpha,\beta}(X:Y)_{P_{XY}}$
is non-increasing in $\beta\in(0,\infty)$.
\end{enumerate}
\end{proposition}
In the above proposition, we do not include the situation $\beta=1$, since $\widetilde{I}_{1,\beta}(X:Y)_{P_{XY}}$ reduces to the mutual information for any $\beta \in (0, \infty)$.
\begin{IEEEproof}[Proof of Proposition~\ref{prop:mono-beta-I}]
The proof is similar to that of Proposition~\ref{prop:mono-beta}. Let $\infty>a\geq b>0$. When $\alpha\in(1,\infty)$, using Jensen's inequality, we have
\begin{align}
\Big(\sum_{y}P_{Y}(y)\|V_{X|y}\|_{\alpha}^{b}\Big)^{\frac{\alpha}{b(\alpha-1)}} & =\Big(\sum_{y}P_{Y}(y)\|V_{X|y}\|_{\alpha}^{b}\Big)^{\frac{\alpha}{a(\alpha-1)}\cdot\frac{a}{b}}\nonumber\\
& \leq\Big(\sum_{y}P_{Y}(y)\|V_{X|y}\|_{\alpha}^{a}\Big)^{\frac{\alpha}{a(\alpha-1)}},
\end{align}
where $V_{X|y}$ is defined in Eq.~\eqref{eq:vector}. When $\alpha\in(0,1)$, we have
\begin{equation}
\Big(\sum_{y}P_{Y}(y)\|V_{X|y}\|_{\alpha}^{b}\Big)^{\frac{\alpha}{b(\alpha-1)}}  \geq\Big(\sum_{y}P_{Y}(y)\|V_{X|y}\|_{\alpha}^{a}\Big)^{\frac{\alpha}{a(\alpha-1)}}.
\end{equation}
Hence, the monotonicity in $\beta \in (0,\infty)$ holds for each fixed $\alpha \in (0,1) \cup (1,\infty)$. 
The cases $\alpha=0$ and $\infty$ are obtained by taking the limit.
\end{IEEEproof}

The following proposition establishes the data processing inequality of the two-parameter R{\'e}nyi mutual information.
\begin{proposition}[Data Processing Inequality]\label{prop:DPI-I}
Let $X-Y-Z$ be a Markov chain. For $\alpha,\beta\in[1,\infty)$ or $ \alpha,\beta\in(0,1]$, we have
\begin{align}
\widetilde{I}_{\alpha,\beta}(X:Y)_{P_{XY}} & \geq\widetilde{I}_{\alpha,\beta}(X:Z)_{P_{XZ}}, \label{eq:DPI-I-1} 
\end{align}
where $P_{XY}$ and $P_{XZ}$ denote joint probability
distributions of $(X,Y)$ and $(X,Z)$, respectively. 
\end{proposition}
\begin{IEEEproof}
The case $\alpha=1$ follows from the fact that the mutual information satisfies the data processing inequality.
Due to the Markov property, it follows that 
\begin{equation}\label{eq:markov}
P_{X|Z}(x|z)=\sum_{y\in\mathcal{Y}}P_{X|Y}(x|y)P_{Y|Z}(y|z).
\end{equation}
When $\alpha\in(1,\infty)$ and $\beta\in[1,\infty)$, using Minkowski's inequality, we have 
\begin{align}
& \Big(\sum_{x}P_{X}^{1-\alpha}(x)P_{X|Z}^{\alpha}(x|z)\Big)^{\frac{1}{\alpha}}\nonumber \\
= & \Big(\sum_{x}\Big(\sum_{y}P_{X}^{\frac{1-\alpha}{\alpha}}(x)P_{X|Y}(x|y)P_{Y|Z}(y|z)\Big)^{\alpha}\Big)^{\frac{1}{\alpha}}\nonumber \\
\leq & \sum_{y}P_{Y|Z}(y|z)\Big(\sum_{x}P_{X}^{1-\alpha}(x)P_{X|Y}^{\alpha}(x|y)\Big)^{\frac{1}{\alpha}}.
\end{align}
Then, using Jensen's inequality, we obtain
\begin{align}
& \sum_{z}P_{Z}(z)\Big(\sum_{x}P_{X}^{1-\alpha}(x)P_{X|Z}^{\alpha}(x|z)\Big)^{\frac{\beta}{\alpha}}\nonumber \\
\leq & \sum_{z}P_{Z}(z)\Big(\sum_{y}P_{Y|Z}(y|z)\Big(\sum_{x}P_{X}^{1-\alpha}(x)P_{X|Y}^{\alpha}(x|y)\Big)^{\frac{1}{\alpha}}\Big)^{\beta}\nonumber \\
\leq & \sum_{y}P_{Y}(y)\Big(\sum_{x}P_{X}^{1-\alpha}(x)P_{X|Y}^{\alpha}(x|y)\Big)^{\frac{\beta}{\alpha}}.
\end{align}
This directly implies that 
\begin{equation}
\widetilde{I}_{\alpha,\beta}(X:Y)_{P_{XY}}\geq\widetilde{I}_{\alpha,\beta}(X:Z)_{P_{XZ}}.
\end{equation}
The case $\alpha\in(0,1)$ and $\beta\in(0,1]$ can be proved in a similar way. This completes the proof.
\end{IEEEproof}

The following proposition establishes the concavity and convexity of the two-parameter R{\'e}nyi mutual information in input distribution $P_X$ and channel $P_{Y|X}$, respectively.
\begin{proposition}[Concavity and Convexity]\label{prop:convexity-concavity}
Let $P_{XY}\in\mathcal{P}(\mathcal{X}\times\mathcal{Y})$. The following statements hold.
\begin{enumerate}
\item For fixed $P_{Y|X}$, $\widetilde{I}_{\alpha,\beta}(X:Y)_{P_{XY}}$ is concave in $P_X$ for $\alpha \in[1,\infty)$ and $\beta\in(0,1]$.
\item For fixed $P_X$, $\widetilde{I}_{\alpha,\beta}(X:Y)_{P_{XY}}$ is convex in $P_{Y|X}$ for $\alpha,\beta \in(0,1]$.
\end{enumerate}
\end{proposition}
\begin{IEEEproof} 
When $\alpha = 1$, this is established in \cite{CoverThomas1991elements}.
We now proceed to Statement 1. From the definition of two-parameter R{\'e}nyi mutual information, we have
\begin{equation}
\widetilde{I}_{\alpha,\beta}(X:Y)_{P_{XY}}=\frac{\alpha}{\beta(\alpha-1)}\log\sum_{y}\Big(\sum_{x}P_X(x)P_{Y|X}(y|x)\Big)^{1-\beta}\Big(\sum_{x'}P_{X}(x')P_{Y|X}^{\alpha}(y|x')\Big)^{\frac{\beta}{\alpha}}.
\end{equation}
Since $\alpha \in(1,\infty)$ and $\beta\in(0,1]$, we have  $1-\beta+\frac{\beta}{\alpha}\in(0,1]$. For any $P_{X},\bar{P}_{X}\in\mathcal{P(X)}$ and $\theta\in[0,1]$, using Lemma~\ref{lem:concave}, we obtain
\begin{align}\label{eq:concave-eq1}
&\Big(\sum_{x}P_{X_\theta}(x)P_{Y|X}(y|x)\Big)^{1-\beta}\Big(\sum_{x'}P_{X_\theta}(x')P_{Y|X}^{\alpha}(y|x')\Big)^{\frac{\beta}{\alpha}} \nonumber\\
\geq& \theta\Big(\sum_{x}P_{X}(x)P_{Y|X}(y|x)\Big)^{1-\beta}\Big(\sum_{x'}P_{X}(x')P_{Y|X}^{\alpha}(y|x')\Big)^{\frac{\beta}{\alpha}}\nonumber\\
&\qquad\qquad+(1-\theta)\Big(\sum_{x}\bar{P}_{X}(x)P_{Y|X}(y|x)\Big)^{1-\beta}\Big(\sum_{x'}\bar{P}_{X}(x')P_{Y|X}^{\alpha}(y|x')\Big)^{\frac{\beta}{\alpha}},
\end{align}
where $P_{X_\theta}=\theta P_{X}+(1-\theta)\bar{P}_{X}$. Since the logarithmic function is concave and $\frac{\alpha}{\beta(\alpha-1)}>0$, Eq.~\eqref{eq:concave-eq1} directly implies the desired result.

Next, we prove Statement 2. For any channels $P_{Y|X},\bar{P}_{Y|X}$, and $\theta\in[0,1]$, using Minkowski's inequality, we have
\begin{equation}\label{eq:concave-eq2}
\Big(\sum_{x}P_{X}(x)P_{Y_\theta|X_\theta}^\alpha(y|x)\Big)^\frac{1}{\alpha}\geq \theta \Big(\sum_{x}P_{X}(x)P_{Y|X}^\alpha(y|x)\Big)^\frac{1}{\alpha}+(1-\theta)\Big(\sum_{x}P_{X}(x) \bar{P}_{Y|X}^\alpha(y|x)\Big)^\frac{1}{\alpha},
\end{equation}
where $P_{Y_\theta|X_\theta}=\theta P_{Y|X}+(1-\theta) \bar{P}_{Y|X}$. From Eq.~\eqref{eq:concave-eq2} and Lemma~\ref{lem:concave}, we derive
\begin{align}
&\Big(\sum_{x}P_{X}(x)P_{Y_\theta|X_\theta}(y|x)\Big)^{1-\beta}\Big(\sum_{x'}P_{X}(x')P_{Y_\theta|X_\theta}^{\alpha}(y|x')\Big)^{\frac{\beta}{\alpha}} \nonumber\\
\geq&\Big(\sum_{x}P_{X}(x)P_{Y_\theta|X_\theta}(y|x)\Big)^{1-\beta}\Big(\theta \Big(\sum_{x}P_{X}(x)P_{Y|X}^\alpha(y|x)\Big)^\frac{1}{\alpha}+(1-\theta)\Big(\sum_{x}P_{X}(x) \bar{P}_{Y|X}^\alpha(y|x)\Big)^\frac{1}{\alpha}\Big)^{\beta} \nonumber\\
\geq&\theta\Big(\sum_{x}P_{X}(x)P_{Y|X}(y|x)\Big)^{1-\beta}\Big(\sum_{x'}P_{X}(x') P_{Y|X}^\alpha(y|x')\Big)^{\frac{\beta}{\alpha}} \nonumber\\
&\qquad\qquad+(1-\theta)\Big(\sum_{x}P_{X}(x)\bar{P}_{Y|X}(y|x)\Big)^{1-\beta}\Big(\sum_{x'}P_{X}(x') \bar{P}_{Y|X}^\alpha(y|x')\Big)^{\frac{\beta}{\alpha}}.
\end{align}
Since the logarithmic function is concave and $\frac{\alpha}{\beta(\alpha-1)}<0$, the desired result follows.
\end{IEEEproof}

In the following Theorem \ref{thm:variational-expression-I}, we give
a variational expression of the two-parameter R{\'e}nyi mutual information,
expressing it as an optimization of linear combination of relative entropies.
\begin{theorem}[Variational Expression]
\label{thm:variational-expression-I}  Let $P_{XY}\in\mathcal{P}(\mathcal{X}\times\mathcal{Y})$. For any $\alpha,\beta\in(0,\infty)$, we have 
\begin{equation}
(1-\alpha)\widetilde{I}_{\alpha,\beta}(X:Y)_{P_{XY}}\!=\!\min_{Q_{XY}\in\mathcal{P}(\mathcal{X}\times\mathcal{Y})}\Big\{\!\frac{\alpha(1\!-\!\beta)}{\beta}D(Q_{Y}\|P_{Y})\!+\!\alpha D(Q_{XY}\|P_{XY})\!+\!(1-\alpha)D(Q_{X|Y}\|P_{X}|Q_{Y})\!\Big\}.\label{eq:variational-I}
\end{equation}
\end{theorem}
\begin{IEEEproof}
The proof is similar to that of Theorem~\ref{thm:variational-expression-H}. When $\alpha=1$, one readily verifies that both sides of Eq.~\eqref{eq:variational-I} vanish. So, we only need to
consider the case $\alpha\in(0,1)\cup(1,\infty)$. 
From the definition of the two-parameter R{\'e}nyi mutual information, we have
\begin{equation}
(1-\alpha)\widetilde{I}_{\alpha,\beta}(X:Y)_{P_{XY}}
=-\frac{\alpha}{\beta}\log\sum_{y}P_{Y}(y)\Big(\sum_{x}P_{X}^{1-\alpha}(x)P_{X|Y}^{\alpha}(x|y)\Big)^{\frac{\beta}{\alpha}}.\label{eq:VE-I-1}
\end{equation}
Define the probability distribution
\begin{equation}
\widetilde{Q}_Y(y)= \frac{P_{Y}(y)\Big(\sum_{x}P_{X}^{1-\alpha}(x)P_{X|Y}^{\alpha}(x|y)\Big)^{\frac{\beta}{\alpha}}}{\sum_{y'}P_{Y}(y')\Big(\sum_{x'}P_{X}^{1-\alpha}(x')P_{X|Y}^{\alpha}(x'|y')\Big)^{\frac{\beta}{\alpha}}},
\end{equation} 
and for each $y\in\mathcal{Y}$ define the probability distribution
\begin{equation}
\widetilde{Q}_X^{(y)}(x)= \frac{P_X^{1-\alpha}(x)P_{X|Y}^\alpha(x|y)}{\sum_{x'}P_X^{1-\alpha}(x')P_{X|Y}^\alpha(x'|y)}.
\end{equation}
Then, we have
\begin{align}\label{eq:VE-I-2}
&-\log\sum_{y}P_{Y}(y)\Big(\sum_{x}P_{X}^{1-\alpha}(x)P_{X|Y}^{\alpha}(x|y)\Big)^{\frac{\beta}{\alpha}}\nonumber \\
=&\min_{Q_{Y}\in\mathcal{P}( \mathcal{Y})} \Big\{D(Q_{Y}\|\widetilde{Q}_{Y})-\log\sum_{y}P_{Y}(y)\Big(\sum_{x}P_{X}^{1-\alpha}(x)P_{X|Y}^{\alpha}(x|y)\Big)^{\frac{\beta}{\alpha}}\Big\}\nonumber \\
=&\min_{Q_{Y}\in\mathcal{P}( \mathcal{Y})}\Big\{D(Q_{Y}\|P_{Y})-\frac{\beta}{\alpha}\sum_{y}Q_{Y}(y)\log\sum_{x}P_X^{1-\alpha}(x)P_{X|Y}^\alpha(x|y)\Big\}.
\end{align}
Furthermore,
\begin{align}
&-\sum_{y}Q_{Y}(y)\log\sum_{x}P_X^{1-\alpha}(x)P_{X|Y}^\alpha(x|y) \nonumber\\
=&\sum_{y}Q_{Y}(y)\Big[\min_{Q_{X}^{(y)}\in\mathcal{P(X)}}\Big(D(Q_{X}^{(y)}\|\widetilde{Q}_X^{(y)})-\log\sum_{x}P_X^{1-\alpha}(x)P_{X|Y}^\alpha(x|y)\Big)\Big]\nonumber \\
=&\sum_{y}Q_{Y}(y)\Big[\min_{Q_{X}^{(y)}\in\mathcal{P(X)}}\Big(D(Q_{X}^{(y)}\|P_{X|y})+(1-\alpha)\sum_{x}Q_{X}^{(y)}(x)\log\frac{P_{X|Y}(x|y)}{P_{X}(x)}\Big)\Big]\nonumber \\
= &\min_{Q_{X|Y}\in\mathcal{P}(\mathcal{X}|\mathcal{Y})}\Big\{D(Q_{X|Y}\|P_{X|Y}|Q_{Y})+(1-\alpha)\sum_{x,y }Q_{XY}(x,y)\log\frac{P_{X|Y}(x|y)}{P_{X}(x)}\Big\},\label{eq:VE-I-3}
\end{align}
where in the last equality we have identified $Q_{X|Y}(\cdot|y)$ with $Q_X^{(y)}$. By Eqs.~\eqref{eq:VE-I-1}, \eqref{eq:VE-I-2} and \eqref{eq:VE-I-3}, we obtain
\begin{align}
&(1-\alpha)\widetilde{I}_{\alpha,\beta}(X:Y)_{P_{XY}}\nonumber\\
= & \min_{Q_{XY}\in\mathcal{P}(\mathcal{X}\times\mathcal{Y})}\Big\{\frac{\alpha}{\beta}D(Q_{Y}\|P_{Y})+D(Q_{X|Y}\|P_{X|Y}|Q_{Y})+(1-\alpha)\sum_{x,y}Q_{XY}(x,y)\log\frac{P_{X|Y}(x|y)}{P_{X}(x)}\Big\}\nonumber \\
= & \min_{Q_{XY}\in\mathcal{P}(\mathcal{X}\times\mathcal{Y})}\Big(\frac{\alpha(1-\beta)}{\beta}D(Q_{Y}\|P_{Y})+\alpha D(Q_{XY}\|P_{XY})+(1-\alpha)D(Q_{X|Y}\|P_{X}|Q_{Y})\Big),\label{eq:variant-expression-I}
\end{align}
where the last equality is obtained from the fact that 
\begin{equation}
\sum_{x,y}Q_{XY}(x,y)\log\frac{P_{X|Y}(x|y)}{P_{X}(x)}=D(Q_{X|Y}\|P_{X}|Q_{Y})-D(Q_{X|Y}\|P_{X|Y}|Q_{Y})
\end{equation}
and 
\begin{equation}
D(Q_{X|Y}\|P_{X|Y}|Q_{Y})=D(Q_{XY}\|P_{XY})-D(Q_{Y}\|P_{Y}).
\end{equation}
\end{IEEEproof}
In \cite{EGI2024sibson}, the authors obtained a variational expression
of Sibson's R{\'e}nyi mutual information, which is the special case
$\beta=1$ of our study. Theorem~\ref{thm:variational-expression-I} shows that $(1-\alpha)\widetilde{I}_{\alpha,\beta}(X:Y)_{P_{XY}}$ is the minimization over a set of functions that are linear in $\alpha\in(0,\infty)$, which directly implies the following corollary.
\begin{corollary}\label{coro:I}
For any $\beta\in(0,\infty)$, the function $(1-\alpha)\widetilde{I}_{\alpha,\beta}(X:Y)_{P_{XY}}$ is concave in $\alpha$ on $(0,\infty)$.
\end{corollary}

\section{Application 1: Strong Converse Exponent of Privacy Amplification}\label{sec:SC-PA}

\subsection{Problem and Main Result}
Let $P_{XY}\in\mathcal{P}(\mathcal{X}\times\mathcal{Y})$ be a joint
probability distribution. We apply a hash function $h:\mathcal{X}\to\mathcal{Z}$ on random variable $X$ to extract randomness
such that the extracted randomness is required to be as private (i.e.,
independent) as possible from $Y$. The distribution induced by the
hash function $h:\mathcal{X}\to\mathcal{Z}$ is given by 
\begin{equation}
\mathcal{R}_{h}(P_{XY})(z,y)=\sum_{x\in h^{-1}(z)}P_{XY}(x,y).
\end{equation}
The goal of privacy amplification is to let the distribution induced
by the hash function approach the ideal distribution $\frac{\mathbbm{1}_{\mathcal{Z}}}{|\mathcal{Z}|}\times P_{Y}$.
We use R{\'e}nyi divergence of order $\beta\in(0,\infty)$ as a measure
of the discrepancy between the real distribution $\mathcal{R}_{h}(P_{XY})$
and ideal distribution $\frac{\mathbbm{1}_{\mathcal{Z}}}{|\mathcal{Z}|}\times P_{Y}$. That is
\begin{equation}
\mathsf{D}_\beta(P_{XY}, h) := D_\beta\Big( \mathcal{R}_h(P_{XY}) \Big\| \frac{\mathbbm{1}_\mathcal{Z}}{|\mathcal{Z}|} \times P_Y \Big).
\end{equation}

In the asymptotic regime, we consider a sequence of hash functions $h_n: \mathcal{X}^n \to \mathcal{Z}_n=\{1,2,\cdots,2^{nR}\}$ applied to $P_{XY}^{\times n}$. Here the non-negative number $R$ is called the extraction rate. Let $\mathcal{A}(R)$ denote the set of such hash function sequences $\{h_n\}_{n=1}^\infty$. The strong converse exponent captures the linear rate at which the divergence grows. It is defined as
\begin{equation}
E_{{\rm pa}}^{(\beta)}(P_{XY},R):=\inf_{\{h_n\}_{n=1}^\infty\in\mathcal{A}(R)} \limsup_{n\to\infty}\frac{1}{n}\mathsf{D}_{\beta}(P_{XY}^{\times n},h_{n}) .\label{eq:def-sce}
\end{equation}

\begin{remark}\label{rem-ap}
The definition of the strong converse exponent based on the order-$\beta$ R{\'e}nyi divergence is equivalent to an alternative formulation involving the order-$\beta$ fidelity. More precisely, in Eq.~\eqref{eq:def-sce}, $\mathsf{D}_{\beta}(P_{XY}^{\times n},h_{n})$
is exactly $-\log\mathsf{F}_{\beta}(P_{XY}^{\times n},h_{n})$, where 
\begin{equation}
\mathsf{F}_{\beta}(P_{XY}^{\times n}, h_n) := F_{\beta}\Big(\mathcal{R}_{h_n}(P_{XY}^{\times n}), \frac{\mathbbm{1}_{\mathcal{Z}_n}}{|\mathcal{Z}_n|} \times P_Y^{\times n} \Big).
\end{equation}
Here, the order-$\beta$ fidelity quantifies the closeness between the real distribution and ideal distribution. Consequently, the quantity
$E_{{\rm pa}}^{(\beta)}(P_{XY},R)$ describes the slowest exponential
rate at which the order-$\beta$ fidelity vanishes (i.e., privacy
amplification fails). 
\end{remark} 

\begin{theorem} \label{thm:strong-converse-exponent}
For any probability distribution $P_{XY}\in\mathcal{P}(\mathcal{X}\times\mathcal{Y})$
and $R\geq0$, we have 
\begin{equation}
E_{{\rm pa}}^{(\beta)}(P_{XY},R)=\begin{cases}
\begin{array}{ll}
\max\limits _{\alpha\in[\beta,1]}\frac{\beta(1-\alpha)}{\alpha(1-\beta)}\left\{ R-\widetilde{H}_{\alpha,\beta}(X|Y)_{P_{XY}}\right\} , & \beta\in(0,1)\\
\left|R-H_{\beta}(X|Y)_{P_{XY}}\right|^{+}, & \beta\in[1,\infty).
\end{array}\end{cases}\label{eq:pa-result}
\end{equation}
\end{theorem}

In~\cite{HayashiTan2016equivocations}, Hayashi and Tan have already
established the result for $\beta\in[1,2]$, which coincides with
Eq.~\eqref{eq:pa-result}. The converse for all $\beta\ge1$ was also proven by them, as will be shown later in Eq.~\eqref{eq:hayashi-tan}.

We also point out that, in an independent work~\cite{BertaYao2025strong}, Berta and Yao have derived the strong converse exponent for privacy amplification with respect to the purified distance. Their Theorem 8 is closely related to our Theorem~\ref{thm:strong-converse-exponent} with $\beta=\frac{1}{2}$.

\subsection{Proof of the Achievability Part: $\beta\in(0,1)$}
In this subsection, we prove the achievability part of Theorem \ref{thm:strong-converse-exponent}
for $\beta\in(0,1)$. 
\begin{proposition} \label{thm:sc-achievability-part}Let
$\beta\in(0,1)$. For any probability distribution $P_{XY}\in\mathcal{P}(\mathcal{X}\times\mathcal{Y})$
and $R\geq0$, we have 
\begin{equation}
E_{{\rm pa}}^{(\beta)}(P_{XY},R)\leq \max_{\alpha\in[\beta,1]}\frac{\beta(1-\alpha)}{\alpha(1-\beta)}\Big\{ R-\widetilde{H}_{\alpha,\beta}(X|Y)_{P_{XY}}\Big\}.
\end{equation}
\end{proposition}
To prove Proposition \ref{thm:sc-achievability-part}, we define 
\begin{align*}
G_{\beta}^{(1)}(P_{XY},R) & :=\inf_{Q_{XY}\in\mathcal{F}_{1}}\Big\{ D(Q_{Y}\|P_{Y})+\frac{\beta}{1-\beta}D(Q_{XY}\|P_{XY})\Big\} \\
G_{\beta}^{(2)}(P_{XY},R) & :=\inf_{Q_{XY}\in\mathcal{F}_{2}}\Big\{ D(Q_{Y}\|P_{Y})+\frac{\beta}{1-\beta}D(Q_{XY}\|P_{XY})+R-H(X|Y)_{Q_{XY}}\Big\} ,
\end{align*}
where 
\begin{align*}
\mathcal{F}_{1} & :=\left\{ Q_{XY}:Q_{XY}\in\mathcal{P}(\mathcal{X}\times \mathcal{Y}),R<H(X|Y)_{Q_{XY}}\right\} \\
\mathcal{F}_{2} & :=\left\{ Q_{XY}:Q_{XY}\in\mathcal{P}(\mathcal{X}\times \mathcal{Y}),R\geq H(X|Y)_{Q_{XY}}\right\} .
\end{align*}
By Theorem~\ref{thm:variational-expression-H}, we have the following lemma. 
\begin{lemma}\label{lem:var-H}
Let $\beta\in(0,1)$, $R\geq0$ and $P_{XY}\in\mathcal{P}(\mathcal{X}\times\mathcal{Y})$.
It holds that
\begin{align}
\max_{\alpha\in[\beta,1]}\frac{\beta\left(1-\alpha\right)}{\alpha\left(1-\beta\right)}\left\{ R-\widetilde{H}_{\alpha,\beta}(X|Y)_{P_{XY}}\right\} =\min\left\{ G_{\beta}^{(1)}(P_{XY},R),G_{\beta}^{(2)}(P_{XY},R)\right\}.
\end{align}
\end{lemma}
\begin{IEEEproof}
From Theorem \ref{thm:variational-expression-H}, we get
\begin{align}
& \max_{\alpha\in[\beta,1]}\frac{\beta(1-\alpha)}{\alpha(1-\beta)}\Big\{ R-\widetilde{H}_{\alpha,\beta}(X|Y)_{P_{XY}}\Big\} \nonumber \\
= & \max_{\alpha\in[\beta,1]}\min_{Q_{XY}\in\mathcal{P}(\mathcal{X}\times\mathcal{Y})}\Big\{\frac{\beta(1\!-\!\alpha)}{\alpha(1\!-\!\beta)}R\!+\!\frac{\beta}{\alpha(1\!-\!\beta)}\Big(\frac{\alpha(1\!-\!\beta)}{\beta}D(Q_{Y}\|P_{Y})\!+\!\alpha D(Q_{XY}\|P_{XY})\!+\!(\alpha\!-\!1)H(X|Y)_{Q_{XY}}\Big)\Big\}\nonumber \\
\overset{(a)}{=} & \max_{\lambda\in[0,1]}\min_{Q_{XY}\in\mathcal{P}(\mathcal{X}\times\mathcal{Y})}\Big\{D(Q_{Y}\|P_{Y})+\frac{\beta}{1-\beta}D(Q_{XY}\|P_{XY})+\lambda\Big(R-H(X|Y)_{Q_{XY}}\Big)\Big\}\nonumber \\
\overset{(b)}{=} & \min_{Q_{XY}\in\mathcal{P}(\mathcal{X}\times\mathcal{Y})}\max_{\lambda\in[0,1]}\Big\{D(Q_{Y}\|P_{Y})+\frac{\beta}{1-\beta}D(Q_{XY}\|P_{XY})+\lambda\Big(R-H(X|Y)_{Q_{XY}}\Big)\Big\}\nonumber \\
= & \min_{Q_{XY}\in\mathcal{P}(\mathcal{X}\times\mathcal{Y})}\Big\{D(Q_{Y}\|P_{Y})+\frac{\beta}{1-\beta}D(Q_{XY}\|P_{XY})+|R-H(X|Y)_{Q_{XY}}|^{+}\Big\},
\end{align}
where $(a)$ is by setting $\frac{\beta(1-\alpha)}{\alpha(1-\beta)}=\lambda$
and $(b)$ comes from Sion's minimax theorem. To see that Sion's minimax
theorem applies here, we have (i) the function $\lambda\mapsto\lambda\left(R-H(X|Y)_{Q_{XY}}\right)$
is linear and continuous, and (ii) the function $Q_{XY}\mapsto D(Q_{Y}\|P_{Y})+\frac{\beta}{1-\beta}D(Q_{XY}\|P_{XY})+\lambda\left(R-H(X|Y)_{Q_{XY}}\right)$ is convex and lower semi-continuous. The desired result follows.
\end{IEEEproof}

\begin{IEEEproof}[Proof of Proposition~\ref{thm:sc-achievability-part}]
This is accomplished by the combination of Lemma~\ref{lem:var-H} and the following Lemmas~\ref{lem:sc-G1} and \ref{lem:sc-G2}.
\end{IEEEproof}

\begin{lemma}
\label{lem:sc-G1}Let $\beta\in(0,1)$. For any probability distribution
$P_{XY}\in\mathcal{P}(\mathcal{X}\times\mathcal{Y})$ and $R\geq0$,
we have 
\begin{equation}
E_{{\rm pa}}^{(\beta)}(P_{XY},R)\leq G_{\beta}^{(1)}(P_{XY},R).
\end{equation}
\end{lemma}
\begin{IEEEproof}
By the definition of $G_{\beta}^{(1)}(P_{XY},R)$, for any $\epsilon\geq0$,
there exists a joint distribution $Q_{XY}\in\mathcal{P}(\mathcal{X}\times\mathcal{Y})$
such that 
\begin{align}
R & <H(X|Y)_{Q_{XY}},\label{eq:sc-achievable-rate}\\
D(Q_{Y}\|P_{Y})+\frac{\beta}{1-\beta}D(Q_{XY}\|P_{XY}) & \leq G_{\beta}^{(1)}(P_{XY},R)+\epsilon.\label{eq:sc-achi-G1}
\end{align}
Csisz{\'a}r \cite{Csiszar1996almost} actually established that for any $R$ satisfying Eq.~\eqref{eq:sc-achievable-rate}, there exists a sequence of hash functions $\{ h_{n}:\mathcal{X}^{\times n}\to\mathcal{Z}_{n}=\{ 1,2,\cdots,2^{nR}\}\} _{n\in\mathbb{N}}$ such that 
\begin{equation}\label{eq:achi-relative}
\lim_{n\to\infty}D\Big(\mathcal{R}_{h_{n}}(Q_{XY}^{\times n})\Big\|\frac{\mathbbm{1}_{\mathcal{Z}_{n}}}{|\mathcal{Z}_{n}|}\times Q_{Y}^{\times n}\Big)=0.
\end{equation}
By the variational expression for the R{\'e}nyi divergence and the data processing inequality (Lemma~\ref{lem:RenyiD-properties}), we have that
\begin{align}\label{eq:achi-relative-beta}
& D_{\beta} \left(\mathcal{R}_{h_n}(P_{XY}^{\times n})\Big\| \frac{\mathbbm{1}_{\mathcal{Z}_n}}{|\mathcal{Z}_n|} \times P_Y^{\times n} \right) \nonumber \\
\leq & D \left(\mathcal{R}_{h_n}(Q_{XY}^{\times n})\Big\| \frac{\mathbbm{1}_{\mathcal{Z}_n}}{|\mathcal{Z}_n|} \times P_Y^{\times n} \right) 
+ \frac{\beta}{1 - \beta} D \left( \mathcal{R}_{h_n}(Q_{XY}^{\times n})\big\| \mathcal{R}_{h_n}(P_{XY}^{\times n}) \right) \nonumber \\
\leq & D \left(\mathcal{R}_{h_n}(Q_{XY}^{\times n})\Big\| \frac{\mathbbm{1}_{\mathcal{Z}_n}}{|\mathcal{Z}_n|} \times Q_Y^{\times n} \right)  + D(Q_Y^{\times n}\|P_Y^{\times n})+\frac{\beta}{1 - \beta} D(Q_{XY}^{\times n} \| P_{XY}^{\times n}) \nonumber \\
= &D \left(\mathcal{R}_{h_n}(Q_{XY}^{\times n})\Big\| \frac{\mathbbm{1}_{\mathcal{Z}_n}}{|\mathcal{Z}_n|} \times Q_Y^{\times n} \right)+ n D(Q_Y \| P_Y)  + \frac{n\beta}{1 - \beta} D(Q_{XY} \| P_{XY}).
\end{align}
Combining Eqs.\eqref{eq:achi-relative} and \eqref{eq:achi-relative-beta} yields
\begin{equation}
\limsup_{n \to \infty} \frac{1}{n} D_{\beta} \Big( \mathcal{R}_{h_n}(P_{XY}^{\times n})\Big\|\frac{\mathbbm{1}_{\mathcal{Z}_n}}{|\mathcal{Z}_n|} \times P_Y^{\times n} \Big) 
\leq D(Q_Y \| P_Y) + \frac{\beta}{1 - \beta} D(Q_{XY} \| P_{XY}). \label{eq:sc-F-upper-bound}
\end{equation}
From Eq.~\eqref{eq:sc-F-upper-bound} and the definition of $E_{\rm{pa}}^{(\beta)}(P_{XY}, R)$, we get
\begin{equation}
E_{\rm{pa}}^{(\beta)}(P_{XY}, R) 
\leq D(Q_Y \| P_Y) + \frac{\beta}{1 - \beta} D(Q_{XY} \| P_{XY}) 
\leq G_{\beta}^{(1)}(P_{XY},R)+\epsilon,
\end{equation}
where the second inequality follows from Eq.~\eqref{eq:sc-achi-G1}.  
Since $\epsilon > 0$ is arbitrary, we conclude the proof by letting $\epsilon \to 0$.
\end{IEEEproof}
\begin{lemma}
\label{lem:sc-G2}Let $\beta\in(0,1)$. For any probability distribution
$P_{XY}\in\mathcal{P}(\mathcal{X}\times\mathcal{Y})$ and $R\geq0$,
we have 
\begin{equation}
E_{{\rm pa}}^{(\beta)}(P_{XY},R)\leq G_{\beta}^{(2)}(P_{XY},R).
\end{equation}
\end{lemma}
\begin{IEEEproof}
By the definition of $G_{\beta}^{(2)}(P_{XY},R)$, there
exists a joint distribution $Q_{XY}\in\mathcal{P}(\mathcal{X}\times\mathcal{Y})$
such that 
\begin{align}
G_{\beta}^{(2)}(P_{XY},R) & =D(Q_{Y}\|P_{Y})+\frac{\beta}{1-\beta}D(Q_{XY}\|P_{XY})+R-H(X|Y)_{Q_{XY}},\\
H(X|Y)_{Q_{XY}} & \leq R.
\end{align}
For any $\epsilon>0$, Let $R':=H(X|Y)_{Q_{XY}}-\epsilon$. Lemma \ref{lem:sc-G1}
shows that there exists a sequence of hash functions $\{ h_{n}^{'}:\mathcal{X}^{\times n}\to\mathcal{Z}_{n}^{'}=\{ 1,2,\cdots,2^{nR'}\}\} _{n\in\mathbb{N}}$
such that 
\begin{equation}
\limsup_{n\to\infty}\frac{1}{n}D_{\beta}\Big(\mathcal{R}_{h_{n}^{'}}(P_{XY}^{\times n})\Big\|\frac{\mathbbm{1}_{\mathcal{Z}_{n}^{'}}}{|\mathcal{Z}_{n}^{'}|}\times P_{Y}^{\times n}\Big)\leq D(Q_{Y}\|P_{Y})+\frac{\beta}{1-\beta}D(Q_{XY}\|P_{XY}).\label{eq:sc-fidelity-G2-initial}
\end{equation}
We transform $\{h_n'\}_{n \in \mathbb{N}}$ into a new sequence of hash functions
\begin{equation}
\{h_n : \mathcal{X}^{\times n} \rightarrow \mathcal{Z}_n = \{1, 2, \ldots, 2^{nR}\}\}_{n \in \mathbb{N}},
\end{equation} 
by expanding the output ranges to accommodate larger amount of extracted
randomness (noting that $R>R'$), while keeping the functions themselves
unchanged, i.e., $h_{n}=h_{n}'$. Under this construction, we obtain
the following result.  
\begin{align}
& D_{\beta}\Big(\mathcal{R}_{h_{n}}(P_{XY}^{\times n})\Big\|\frac{\mathbbm{1}_{\mathcal{Z}_{n}}}{|\mathcal{Z}_{n}|}\times P_{Y}^{\times n}\Big)\nonumber \\
= & D_{\beta}\Big(\mathcal{R}_{h_{n}^{'}}(P_{XY}^{\times n})\Big\|\frac{\mathbbm{1}_{\mathcal{Z}_{n}^{'}}}{|\mathcal{Z}_{n}|}\times P_{Y}^{\times n}\Big)\nonumber \\
= & D_{\beta}\Big(\mathcal{R}_{h_{n}^{'}}(P_{XY}^{\times n})\Big\|\frac{\mathbbm{1}_{\mathcal{Z}_{n}^{'}}}{|\mathcal{Z}_{n}'|}\times P_{Y}^{\times n}\Big)+\log\frac{|\mathcal{Z}_{n}|}{|\mathcal{Z}_{n}'|}\nonumber \\
= & D_{\beta}\Big(\mathcal{R}_{h_{n}^{'}}(P_{XY}^{\times n})\Big\|\frac{\mathbbm{1}_{\mathcal{Z}_{n}^{'}}}{|\mathcal{Z}_{n}'|}\times P_{Y}^{\times n}\Big)+n(R-R').\label{eq:sc-fidelity-G2}
\end{align}
Combining Eqs.~\eqref{eq:sc-fidelity-G2-initial} and \eqref{eq:sc-fidelity-G2}, we obtain
\begin{align}
E_{\rm{pa}}^{(\beta)}(P_{XY},R) & \leq\limsup_{n\to\infty}\frac{1}{n}D_{\beta}\Big(\mathcal{R}_{h_{n}}(P_{XY}^{\times n})\Big\|\frac{\mathbbm{1}_{\mathcal{Z}_{n}}}{|\mathcal{Z}_{n}|}\times P_{Y}^{\times n}\Big)\nonumber \\
& \leq D(Q_{Y}\|P_{Y})+\frac{\beta}{1-\beta}D(Q_{XY}\|P_{XY})+R-R'\nonumber \\
& = G_{\beta}^{(2)}(P_{XY},R)+\epsilon.\label{eq:sc-G2-final}
\end{align}
Since Eq.~\eqref{eq:sc-G2-final} holds for any $\epsilon>0$,
the conclusion follows by taking the limit $\epsilon\to0$.
\end{IEEEproof}
\subsection{Proof of the Achievability Part: $\beta\in[1,\infty)$}
Having addressed the case $\beta\in(0,1)$, we now turn to the other case $\beta\in[1,\infty)$. To obtain the claimed result, we will employ the following Lemma~\ref{lem:k-universal} given in \cite{PathegamaBarg2024renyi}. Before stating the lemma, we introduce two essential definitions.

A family of hash functions $\mathcal{H}=\{h:\mathcal{X}\to\mathcal{Z}\}$
is called $k$-universal \cite{BBCM2002generalized}, if for all distinct
elements $x_{1},x_{2},\cdots,x_{k}\in\mathcal{X}$, we have 
\begin{equation}
\mathbb{P}_{\mathcal{H}}\left(h\in\mathcal{H}:h(x_{1})=h(x_{2})=\cdots=h(x_{k})\right)\leq|\mathcal{Z}|^{1-k},
\end{equation}
where $\mathbb{P}_{\mathcal{H}}$ denotes the counting probability
measure on the family $\mathcal{H}$. The family $\mathcal{H}$ is
called $k^{*}$-universal \cite{PathegamaBarg2024renyi} if it is
$l$-universal for all $l\in\{2,3,\cdots,k\}$. 
\begin{lemma}[\cite{PathegamaBarg2024renyi}]\label{lem:k-universal}
Let $k\in\{2,3,\cdots\}$ and $\alpha\in(1,k]$. Let $\mathcal{H}=\left\{ h:\mathcal{X}\to\mathcal{Z}\right\} $
be a $k^{*}$-universal family of hash functions and $P_{XY}\in\mathcal{P}(\mathcal{X}\times\mathcal{Y})$
be a probability distribution. Then, 
\begin{align}
&\mathbbm{E}_{h\sim \mathbb{P}_{\mathcal{H}}}\exp\Big\{ (\alpha-1)D_{\alpha}\Big(\mathcal{R}_{h}(P_{XY})\Big\|\frac{\mathbbm{1}_{\mathcal{Z}}}{|\mathcal{Z}|}\times P_{Y}\Big)\Big\} \nonumber\\
\leq& \sum_{l=1}^{\left\lceil \alpha\right\rceil -1}l\left\{\!\!\! \begin{array}{c}
\left\lceil \alpha\right\rceil\! -\!1\\
l
\end{array}\!\!\!\right\} \exp\{(\alpha-l)(\log|\mathcal{Z}|\!-\!H_{\alpha}(X|Y)_{P_{XY}})\} \nonumber\\
&\qquad+\sum_{l=1}^{\left\lceil \alpha\right\rceil -1}\left\{\!\!\! \begin{array}{c}
\left\lceil \alpha\right\rceil\! -\!1\\
l-1
\end{array}\!\!\!\right\}\exp\{(\left\lceil \alpha\right\rceil \! -\!l)(\log|\mathcal{Z}|\!-\!H_{\alpha}(X|Y)_{P_{XY}})\}\!+\!1,
\end{align}
where $\genfrac{\{}{\}}{0pt}{}{\scriptstyle i}{\scriptstyle j}$
denotes the Stirling number of the second kind, which equals the number of ways to partition a set of $i$ elements into $j$ nonempty subsets.
\end{lemma}
This one-shot bound directly implies the following asymptotic result. 
\begin{lemma}
\label{lem:achievability-k} Let $R\geq0$ and $\alpha>1$. For each $n\in\mathbb{N}$, let $\mathcal{H}_{n}=\{ h_{n}:\mathcal{X}^{\times n}\to\mathcal{Z}_{n}=\{1,2,\cdots,2^{nR}\}\} $
be a family of $k^{*}$-universal hash functions with $k=\min\{\left\lceil \alpha\right\rceil,|\mathcal{X}|^n \}$. For any probability distribution $P_{XY}\in\mathcal{P}(\mathcal{X}\times\mathcal{Y})$, there exists $h_{n}\in \mathcal{H}_{n}$
such that
\begin{equation}
D_{\alpha}\Big(\mathcal{R}_{h_{n}}(P_{XY}^{\times n})\Big\|\frac{\mathbbm{1}_{\mathcal{Z}_{n}}}{|\mathcal{Z}_{n}|}\times P_{Y}^{\times n}\Big)\overset{.}{\leq}\max_{l\in\{\left\lceil \alpha\right\rceil -1\}\cup\{\alpha-[\left\lceil \alpha\right\rceil -1]\}}\Gamma(l),
\end{equation}
where $\Gamma(l):=\exp\{l(R-H_{\alpha}(X|Y)_{P_{XY}})\}$. 
\end{lemma}
By applying the proof technique used in Lemma \ref{lem:sc-G2} again,
we derive the achievability part of the strong converse exponent for $\beta\in[1,\infty)$. 
\begin{proposition}
Let $\beta\in[1,\infty)$. For any probability distribution $P_{XY}$
and $R\geq0$, we have 
\begin{equation}
E_{\rm{pa}}^{(\beta)}(P_{XY},R)\leq\left|R-H_{\beta}(X|Y)_{P_{XY}}\right|^{+}.
\end{equation}
\end{proposition}
\begin{IEEEproof}
For the case $R<H_{\beta}(X|Y)_{P_{XY}}$, from Lemma \ref{lem:achievability-k},
we directly get
\begin{equation}
E_{\rm{pa}}^{(\beta)}(P_{XY},R)\leq0.
\end{equation}

Consider the other case $R\geq H_{\beta}(X|Y)_{P_{XY}}$. Let $R':=H_{\beta}(X|Y)_{P_{XY}}-\epsilon$ with $\epsilon>0$ being arbitrary. Lemma \ref{lem:achievability-k} shows that there is
a sequence of hash functions $\{ h_{n}^{'}:\mathcal{X}^{\times n}\to\mathcal{Z}_{n}^{'}=\{ 1,\cdots,2^{nR'}\} \} _{n\in\mathbb{N}}$
such that 
\begin{equation}
\limsup_{n\to\infty}\frac{1}{n}D_{\beta}\Big(\mathcal{R}_{h_{n}^{'}}(P_{XY}^{\times n})\Big\|\frac{\mathbbm{1}_{\mathcal{Z}_{n}^{'}}}{|\mathcal{Z}_{n}^{'}|}\times P_{Y}^{\times n}\Big)\leq0.\label{eq:sc-beta-1}
\end{equation}
We transform $\{h_n'\}_{n \in \mathbb{N}}$ into a new sequence of hash functions  
\begin{equation}
\{h_n : \mathcal{X}^{\times n} \rightarrow \mathcal{Z}_n = \{1, 2, \ldots, 2^{nR}\}\}_{n \in \mathbb{N}},
\end{equation}  
by extending the output ranges to accommodate a larger amount of extracted randomness,  
while keeping the functions themselves unchanged, i.e., $h_n = h_n'$. Then we have the following result.

\begin{align}
& D_{\beta}\Big(\mathcal{R}_{h_{n}}(P_{XY}^{\times n})\Big\|\frac{\mathbbm{1}_{\mathcal{Z}_{n}}}{|\mathcal{Z}_{n}|}\times P_{Y}^{\times n}\Big)\nonumber \\
= & D_{\beta}\Big(\mathcal{R}_{h_{n}^{'}}(P_{XY}^{\times n})\Big\|\frac{\mathbbm{1}_{\mathcal{Z}_{n}^{'}}}{|\mathcal{Z}_{n}|}\times P_{Y}^{\times n}\Big)\nonumber \\
= & D_{\beta}\Big(\mathcal{R}_{h_{n}^{'}}(P_{XY}^{\times n})\Big\|\frac{\mathbbm{1}_{\mathcal{Z}_{n}^{'}}}{|\mathcal{Z}_{n}'|}\times P_{Y}^{\times n}\Big)+\log\frac{|\mathcal{Z}_{n}|}{|\mathcal{Z}_{n}'|}\nonumber \\
= & D_{\beta}\Big(\mathcal{R}_{h_{n}^{'}}(P_{XY}^{\times n})\Big\|\frac{\mathbbm{1}_{\mathcal{Z}_{n}^{'}}}{|\mathcal{Z}_{n}'|}\times P_{Y}^{\times n}\Big)+n(R-R').\label{eq:sc-beta-2}
\end{align}
Combining Eqs.~\eqref{eq:sc-beta-1} and \eqref{eq:sc-beta-2}, we obtain
\begin{align}
E_{\rm{pa}}^{(\beta)}(P_{XY},R) & \leq\limsup_{n\to\infty}\frac{1}{n}D_{\beta}\Big(\mathcal{R}_{h_{n}}(P_{XY}^{\times n})\Big\|\frac{\mathbbm{1}_{\mathcal{Z}_{n}}}{|\mathcal{Z}_{n}|}\times P_{Y}^{\times n}\Big)\nonumber \\
& \leq R-R'\nonumber \\
& = R-H_{\beta}(X|Y)_{P_{XY}}+\epsilon.\label{eq:sc-G2-final-1}
\end{align}
Since Eq.~\eqref{eq:sc-G2-final-1} holds for any $\epsilon > 0$,  
letting $\epsilon \to 0$ completes the proof.
\end{IEEEproof}

\subsection{Proof of the Optimality Part}
In this subsection, we establish the optimality part of Theorem \ref{thm:strong-converse-exponent}.

\begin{lemma}
\label{lem:fidelity-upper-bound} Let $P_{XY}\in\mathcal{P}(\mathcal{X}\times\mathcal{Y})$ be a probability
distribution. For $\beta\in(0,1)$ and $\alpha\in[\beta,1)$, it holds that
\begin{equation}
\frac{\alpha(1-\beta)}{\beta(1-\alpha)}D_{\beta}\Big(P_{XY}\Big\|\frac{\mathbbm{1}_\mathcal{X}}{|\mathcal{X}|}\times P_{Y}\Big)\geq\log|\mathcal{X}|-\widetilde{H}_{\alpha,\beta}(X|Y)_{P_{XY}}.
\end{equation}
\end{lemma}

\begin{IEEEproof}
When $\alpha\geq\beta$, Hölder's inequality implies that 
\begin{align}
\sum_{x}P_{X|Y}^{\beta}(x|y)= & \sum_{x}1\cdot P_{X|Y}^{\beta}(x|y)\nonumber \\
\leq & |\mathcal{X}|^{\frac{\alpha-\beta}{\alpha}}\Big(\sum_{x}P_{X|Y}^{\alpha}(x|y)\Big)^{\frac{\beta}{\alpha}}.
\end{align}
Using this relation, we obtain that, for $\alpha\in[\beta,1)$, 
\begin{align}
& \frac{\alpha(1-\beta)}{\beta(1-\alpha)}D_{\beta}\left(P_{XY}\Big\|\frac{\mathbbm{1}_\mathcal{X}}{|\mathcal{X}|}\times P_{Y}\right)\nonumber \\
= & \frac{-\alpha}{\beta(1-\alpha)}\log\Big(|\mathcal{X}|^{\beta-1}\sum_{y}P_{Y}(y)\sum_{x}P_{X|Y}^{\beta}(x|y)\Big)\nonumber \\
\geq & \frac{-\alpha}{\beta(1-\alpha)}\log\Big(|\mathcal{X}|^{\frac{(\alpha-1)\beta}{\alpha}}\sum_{y}P_{Y}(y)\Big(\sum_{x}P_{X|Y}^{\alpha}(x|y)\Big)^{\frac{\beta}{\alpha}}\Big)\nonumber \\
= & \log|\mathcal{X}|-\widetilde{H}_{\alpha,\beta}(X|Y)_{P_{XY}}.
\end{align}
\end{IEEEproof}

With Lemma~\ref{lem:fidelity-upper-bound}, we are able to prove the the optimality part of Theorem \ref{thm:strong-converse-exponent} for $\beta\in(0,1)$.
\begin{proposition} \label{prop:sce-opt}
Let $\beta\in(0,1)$. For any probability distribution $P_{XY}\in\mathcal{P}(\mathcal{X}\times\mathcal{Y})$ and $R\geq0$, we have 
\begin{equation}
E_{{\rm pa}}^{(\beta)}(P_{XY},R)\geq \max_{\alpha\in[\beta,1]}\frac{\beta(1-\alpha)}{\alpha(1-\beta)}\Big\{ R-\widetilde{H}_{\alpha,\beta}(X|Y)_{P_{XY}}\Big\}.
\end{equation}
\end{proposition}

\begin{IEEEproof}
Let $h_n : \mathcal{X}^{\times n} \rightarrow \mathcal{Z}_n = \{1, 2, \ldots, 2^{nR}\}$ be a sequence of hash functions. For any $\beta\in(0,1)$ and $\alpha\in[\beta,1)$, we have 
\begin{align}
& \frac{\alpha(1-\beta)}{\beta(1-\alpha)}D_{\beta}\Big(\mathcal{R}_{h_{n}}(P_{XY}^{\times n})\Big\|\frac{\mathbbm{1}_{\mathcal{Z}_{n}}}{|\mathcal{Z}_{n}|}\times P_{Y}^{\times n}\Big)\nonumber \\
\geq & \log|\mathcal{Z}_{n}|-\widetilde{H}_{\alpha,\beta}(Z_{n}|Y^{n})_{\mathcal{R}_{h_{n}}(P_{XY}^{\times n})}\nonumber \\
\geq & \log|\mathcal{Z}_{n}|-\widetilde{H}_{\alpha,\beta}(X^{n}|Y^{n})_{P_{XY}^{\times n}}\nonumber \\
= & nR-n\widetilde{H}_{\alpha,\beta}(X|Y)_{P_{XY}},
\end{align}
where the first inequality follows from Lemma \ref{lem:fidelity-upper-bound}, the second inequality comes from Proposition~\ref{prop:entropy-mono} and the last equality is because $\widetilde{H}_{\alpha,\beta}(X|Y)_{P_{XY}}$ is additive (Proposition~\ref{prop:additivity}).
So,
\begin{align}
& \liminf_{n\to\infty}\frac{1}{n}D_{\beta}\Big(\mathcal{R}_{h_{n}}(P_{XY}^{\times n})\Big\|\frac{\mathbbm{1}_{\mathcal{Z}_{n}}}{|\mathcal{Z}_{n}|}\times P_{Y}^{\times n}\Big)\nonumber \\
\geq & \max_{\beta\leq\alpha\leq1}\frac{\beta\left(1-\alpha\right)}{\alpha\left(1-\beta\right)}\left\{ R-\widetilde{H}_{\alpha,\beta}(X|Y)_{P_{XY}}\right\} .
\end{align}
By the definition of $E_{{\rm sc}}^{(\beta)}(P_{XY},R)$, we conclude
the proof.
\end{IEEEproof}

For the case $\beta\geq1$, Hayashi and Tan~\cite{HayashiTan2016equivocations}
have shown that for any sequence of hash functions $h_n : \mathcal{X}^{\times n} \rightarrow \mathcal{Z}_n = \{1, 2, \ldots, 2^{nR}\}$,
\begin{equation}\label{eq:hayashi-tan}
\liminf_{n\to\infty}\frac{1}{n}D_{\beta}\Big(\mathcal{R}_{h_{n}}(P_{XY}^{\times n})\Big\|\frac{\mathbbm{1}_{\mathcal{Z}_{n}}}{|\mathcal{Z}_{n}|}\times P_{Y}^{\times n}\Big)\geq\left|R-H_{\beta}(X|Y)_{P_{XY}}\right|^{+},
\end{equation}
which leads to 
\begin{equation}
E_{\rm{pa}}^{(\beta)}(P_{XY},R)\geq\left|R-H_{\beta}(X|Y)_{P_{XY}}\right|^{+}.
\end{equation}.

\section{Application 2: Strong Converse Exponent of Soft Covering}\label{sec:SC-SC}

\subsection{Problem and Main Result}

Let $P_{Y|X}$ represent a discrete memoryless channel from alphabet $\mathcal{X}$ to $\mathcal{Y}$ and $P_{X}\in\mathcal{P(X)}$ be an input distribution. Then the output distribution is
\begin{equation}
P_{Y}(y)=\sum_{x\in\mathcal{X}}P_{X}(x)P_{Y|X}(y|x).
\end{equation}
The goal of soft covering is to approximate this marginal distribution
at the channel output, given access to the channel $P_{Y|X}$ and the
ability to sample from the input distribution $P_{X}$.

To this end, we consider a random code $\mathcal{C}=\{X(m)\}_{m=1}^{M}$
of size $M$, where each codeword $X(m)$ is independently drawn from $P_{X}$. The average output distribution induced by the code
$\mathcal{C}$ is then given by
\begin{equation}
P_{Y|\mathcal{C}}(y)=\sum_{m=1}^M \frac{1}{M} P_{Y|X}(y| X(m)). \label{eq:}
\end{equation}
Note that here $P_{Y|\mathcal{C}}(y)$ is a random variable, and for a specific realization of the code $\mathcal{C}=\{x(m)\}_{m=1}^M$, it takes the value $\sum_{m=1}^M \frac{1}{M} P_{Y|X}(y| x(m))$.
We use the R{\'e}nyi divergence with order $\beta \in(0,\infty)$ to measure the discrepancy between the code-induced distribution
$P_{Y|\mathcal{C}}$ and the true marginal output distribution
$P_{Y}$. That is,
\begin{align}\label{eq:definition-one-shot}
\mathsf{D}_{\beta}(P_{XY},\mathcal{C}):= D_\beta (P_{Y|\mathcal{C}} || P_Y | P_\mathcal{C})=\left\{
\begin{array}{ll}
\frac{1}{\beta-1}\log\mathbb{E}_{\mathcal{C}}\Big[\sum_{y}P_{Y|\mathcal{C}}^{\beta}(y)P_{Y}^{1-\beta}(y)\Big], &\beta\neq 1 \\
\mathbb{E}_{\mathcal{C}} D(P_{Y|\mathcal{C}}\|P_Y), & \beta=1.
\end{array}\right.
\end{align} 
In the asymptotic setting, there are many independent copies of the
channel, denoted by the product conditional distribution $P_{Y|X}^{\times n}$
with $n\in\mathbb{N}$. Let $\mathcal{C}_{n}=\{X^{n}(m)\}_{m=1}^{2^{nR}}$
be an i.i.d. random code, where each codeword $X^{n}(m)$ is drawn
independently according to $P_{X}^{\times n}$. Here the positive
number $R$ is called the rate. Now the goal is to use the code-induced
distribution $P_{Y^{n}|\mathcal{C}_{n}}$ (defined by the formula
in \eqref{eq:}) to approximate $P_{Y}^{\times n}$. The strong converse
exponent characterizes the linear rate at which the divergence grows,
defined as 
\begin{equation}\label{eq:definition-sc}
E_{{\rm sc}}^{(\beta)}(P_{XY},R):=\lim_{n\to\infty}\frac{1}{n}\mathsf{D}_{\beta}(P_{XY}^{\times n},\mathcal{C}_{n}).
\end{equation}

\begin{remark}
(i) Since $D_\beta(P_{XY}\|\mathcal{C})=-\log F_\beta(P_{\mathcal{C}Y},P_\mathcal{C}\times P_Y)$, the quantity $E_{{\rm sc}}^{(\beta)}(P_{XY},R)$ can also be understood as the rate of exponential convergence to $0$ of the order-$\beta$ fidelity between the code-induced distribution and the distribution $P_{Y}^{\times n}$. 
(ii) In the proof of Theorem~\ref{thm:SC-strong-converse-exponent}, we will confirm that the limit in Eq.~\eqref{eq:definition-sc} does exist.
\end{remark}
\begin{theorem}\label{thm:SC-strong-converse-exponent}
For any probability distribution
$P_{XY}\in\mathcal{P}(\mathcal{X}\times\mathcal{Y})$ and $R\geq0$,
we have 
\begin{equation}
E_{{\rm sc}}^{(\beta)}(P_{XY},R)=
\begin{cases}
\max\limits_{\beta\leq\alpha\leq1}\frac{\beta(1-\alpha)}{\alpha(1-\beta)}\left\{ \widetilde{I}_{\alpha,\beta}(X:Y)_{P_{XY}}-R\right\} , & \beta\in(0,1)\\
\left|I_{\beta}(X:Y)_{P_{XY}}-R\right|^{+}, & \beta\in[1,\infty).
\end{cases}
\end{equation}
\end{theorem}

\subsection{Method of Types and Strong Packing-Covering Lemma for Random I.I.D. Codes}
The method of types~\cite{Csiszar1998method,CsiszarKorner2011information} is a fundamental and widely used tool in information theory. In the following, we introduce several relevant definitions and properties that will be used throughout this paper.

For a sequence $x^{n}:=(x_{1},x_{2},\ldots,x_{n})\in\mathcal{X}^{\times n}$,
we use $T_{x^{n}}(x):=\frac{1}{n}\sum_{i=1}^{n}\boldsymbol{1}{\{x_{i}=x\}}$ to
denote the type of $x^{n}$. The set of all types that the elements of $\mathcal{X}^{\times n}$ can take is denoted by
\begin{equation}
\mathcal{P}_{n}(\mathcal{X}):=\{T_{x^{n}}:x^{n}\in\mathcal{X}^{\times n}\}.
\end{equation}
For sequences $(x^n,y^n)\in\mathcal{X}^{\times n}\times\mathcal{Y}^{\times n}$, a conditional probability distribution $T_{Y|X}:\mathcal{X}\to\mathcal{Y}$ is called the conditional type of $y^n$ given $x^n$ if for any $(x,y)\in\mathcal{X}\times\mathcal{Y}$
\begin{equation}
T_{x^n y^n}(x,y)=T_{Y|X}(y|x)T_{x^n}(x),
\end{equation}
where $T_{x^n y^n}$ and $T_{x^n}$ denote the joint type of $(x^n,y^n)$ and the type of $x^n$, respectively. Given a sequence $ x^n \in \mathcal{X}^{\times n} $, the set of conditional types over $ \mathcal{Y} $ conditioned on $ x^n $ is denoted by  
\begin{equation}
\mathcal{P}_n(\mathcal{Y} | T_{x^n}) := \left\{ T_{Y|X} \in \mathcal{P}(\mathcal{Y} | \mathcal{X}) : T_{x^n} T_{Y|X} \in \mathcal{P}_n(\mathcal{X} \times \mathcal{Y}) \right\}.
\end{equation} 
For any type $T_X \in \mathcal{P}_{n}(\mathcal{X})$, the set of sequences of length $n$ that have type $T_X$ is called the type class of $T_X$, denoted by
\begin{equation}
\mathcal{T}_{T_X} := \{x^{n} \in \mathcal{X}^{\times n} : T_{x^{n}} = T_X \}.
\end{equation} 
For any conditional type $ T_{Y|X} \in \mathcal{P}_n(\mathcal{Y} | T_{X})$, the conditional type class of $T_{Y|X}$ given $x^n$ is
\begin{equation}
\mathcal{T}_{T_{Y|X}}(x^n) := \{y^n\in\mathcal{Y}^{\times n} : T_{Y|X} \text{ is the conditional type of } y^n \text{ given } x^n\}.
\end{equation}
The sizes of $\mathcal{P}_{n}(\mathcal{X})$ and $\mathcal{P}_n(\mathcal{Y} | T_{x^n})$ grow polynomially with $ n $, and satisfy that
\begin{align}
|\mathcal{P}_{n}(\mathcal{X})| \leq& (n+1)^{|\mathcal{X}|}, \label{eq:size-type}\\
|\mathcal{P}_n(\mathcal{Y} | T_{x^n})|\leq& (n+1)^{|\mathcal{X}|\cdot|\mathcal{Y}|}.\label{eq:size-conditional-type}
\end{align} 
The sizes of the type class $ \mathcal{T}_{T_X} $ and the conditional type class $\mathcal{T}_{T_{Y|X}}(x^n)$ can be bounded by
\begin{align}
(n+1)^{-|\mathcal{X}|} \cdot 2^{n H(X)_{T_X}} \leq& |\mathcal{T}_{T_X}| \leq 2^{n H(X)_{T_X}}, \label{eq:type-size}\\
(n+1)^{-|\mathcal{X}|\cdot|\mathcal{Y}|} \cdot 2^{n H(Y|X)_{T_{XY}}} \leq& |\mathcal{T}_{T_{Y|X}}(x^n)| \leq 2^{n H(Y|X)_{T_{XY}}}. \label{eq:conditional-type-class}
\end{align} 
Let $P_X \in \mathcal{P}(\mathcal{X})$ and $T_X\in\mathcal{P}_n(\mathcal{X})$. The probability of the type class $ \mathcal{T}_{T_X} $ under the product distribution $ P_X^{\times n} $ satisfies
\begin{equation}\label{eq:prob}
(n+1)^{-|\mathcal{X}|} \cdot 2^{-n D(T_X \| P_X)} \leq P_X^{\times n}(\mathcal{T}_{T_X}) \leq 2^{-n D(T_X \| P_X)},
\end{equation} 
where  
\begin{equation}
P_X^{\times n}(\mathcal{T}_{T_X}) := \sum_{x^n \in \mathcal{T}_{T_X}} P_X^{\times n}(x^n).
\end{equation} 

Let $R\geq0$ and $\mathcal{C}_{n}:=\{X^{n}(m)\}_{m=1}^{2^{nR}}$ be an i.i.d. random code, where each codeword
$X^{n}(m)$ is drawn independently according to $P_{X}^{\times n}$. For any $\epsilon>0$ and type $T_X\in\mathcal{P}_n(\mathcal{X})$,
define two events on $\mathcal{C}_{n}$ as 
\begin{align}
\mathcal{B}_{1}(\epsilon|T_{X},P_{X})&:=\left\{ \left|\frac{\varphi_{\mathcal{C}_{n}}(T_{X})}{\mathbb{E}\left[\varphi_{\mathcal{C}_{n}}(T_{X})\right]}-1\right|\leq 2^{-n\epsilon}\right\} , \\
\mathcal{B}_{2}(\epsilon|T_{X},P_{X}) & :=\left\{ 0\leq\varphi_{\mathcal{C}_{n}}(T_{X})\leq 2\cdot2^{4n\epsilon}\right\},
\end{align}
where 
\begin{equation}\label{eq:definition-varphi}
\varphi_{\mathcal{C}_{n}}(T_{X}):=\left|\mathcal{T}_{T_{X}}\cap\mathcal{C}_{n}\right|=\sum_{m\in[2^{nR}]}\boldsymbol{1}{\{X^{n}(m)\in\mathcal{T}_{T_{X}}\}}
\end{equation}
is the number of codewords belonging to the type class $\mathcal{T}_{T_{X}}$.
The probability that a sequence $x^n$ drawn i.i.d. from $ P_X^{\times n} $ has type $ T_X $ is  given by $P_X^{\times n}(\mathcal{T}_{T_X})$. From Eq.~\eqref{eq:prob}, we have  $\mathbb{E}\left[\varphi_{\mathcal{C}_{n}}(T_{X})\right]$ satisfies
\begin{equation}
(n+1)^{-|\mathcal{X}|}2^{n(R-D(T_{X}\|P_{X}))}\leq\mathbb{E}\left[\varphi_{\mathcal{C}_{n}}(T_{X})\right]
\leq 2^{n(R-D(T_{X}\|P_{X}))}.\label{eq:size-mean-1}
\end{equation}
If $R-D(T_{X}\|P_{X})<4\epsilon$ , then according to Eq.~\eqref{eq:prob} again, the probability that the random code $\mathcal{C}_n$ contains at least one element of $\mathcal{T}_{T_X}$ satisfies the following inequality.
\begin{align}
\mathbb{P}\{\varphi_{\mathcal{C}_n}(T_X)\geq1\}\leq&1-\left(1-2^{-nD(T_X\|P_X)}\right)^{2^{nR}}\dot{=}1 - e^{-2^{n(R - D(T_X \| P_X))}}\leq 2^{n(R - D(T_X \| P_X))},\label{eq:for-next-sc}\\
\mathbb{P}\{\varphi_{\mathcal{C}_n}(T_X)\geq1\}\geq&1-\left(1-(n+1)^{-|\mathcal{X}|} \cdot 2^{-nD(T_X\|P_X)}\right)^{2^{nR}}\dot{\geq} 2^{n(R - D(T_X \| P_X)-4\epsilon)}.\label{eq:for-next-sc-1}
\end{align} 

For any joint type $T_{XY}\in\mathcal{P}_n(\mathcal{X}\times \mathcal{Y})$, define two events on $\mathcal{C}_{n}$ as 
\begin{align}
\mathcal{B}_{3}(\epsilon|T_{XY},P_{X}) & :=\left\{ \left|\frac{\phi_{\mathcal{C}_{n}}(y^{n})}{\mathbb{E}\left[\phi_{\mathcal{C}_{n}}(y^{n})\right]}-1\right|\leq2^{-n\epsilon},\forall y^{n}\in\mathcal{T}_{T_{Y}}\right\} ,\nonumber \\
\mathcal{B}_{4}(\epsilon|T_{XY},P_{X}) & :=\left\{ 0\leq\phi_{\mathcal{C}_{n}}(y^{n})\leq{2^{5n\epsilon}},\forall y^{n}\in\mathcal{T}_{T_{Y}}\right\} ,
\end{align}
where 
\begin{align}\label{eq:definition-phi}
\phi_{\mathcal{C}_{n}}(y^{n}) & :=\big|\mathcal{T}_{T_{X|Y}}(y^{n})\cap\mathcal{C}_{n}\big|\nonumber\\
& =\sum_{m\in[2^{nR}]}\boldsymbol{1}{\{X^{n}(m)\in\mathcal{T}_{T_{X|Y}}(y^{n})\}}
\end{align}
is the number of codewords belonging to the conditional type class
$\mathcal{T}_{T_{X|Y}}(y^{n})$. It is straightforward to estimate $\mathbb{E}\left[\phi_{\mathcal{C}_{n}}(y^{n})\right]$, which satisfies
\begin{align}
\mathbb{E}\left[\phi_{\mathcal{C}_{n}}(y^{n})\right] & = 2^{n(R-D(T_X\|P_X)-H(X)_{T_X})}\cdot\big|\mathcal{T}_{T_{X|Y}}(y^{n})\big|.
\end{align}
So, by Eq.~\eqref{eq:conditional-type-class} we have
\begin{equation}
(n+1)^{-|\mathcal{X}|\cdot|\mathcal{Y}|}2^{n\left(R-D(T_{X}\|P_{X})-I(X:Y)_{T_{XY}}\right)}\leq \mathbb{E}\left[\phi_{\mathcal{C}_{n}}(y^{n})\right] \leq 2^{n\left(R-D(T_{X}\|P_{X})-I(X:Y)_{T_{XY}}\right)}.
\end{equation}

\begin{lemma}\label{lem:iid}
Let $\epsilon>0$ and $R\geq4\epsilon$ be fixed. Let $\delta_n := \frac{|\mathcal{X}|}{n}\log(n+1)$. Then, the following statements hold.
\begin{enumerate}
\item It holds that
\begin{equation}
\mathbb{P}\left[\mathcal{B}_{1}(\epsilon|T_{X},P_{X})\right]\geq1-2^{-\frac{1}{3}\exp(n(\epsilon-\delta_n))},
\end{equation}
for all types $T_{X}\in\mathcal{P}_n(\mathcal{X})$ such that $R\geq D(T_{X}\|P_{X})+4\epsilon$. 
\item  It holds that
\begin{equation}
\mathbb{P}\left[\mathcal{B}_{2}(\epsilon|T_{X},P_{X})\right]\geq1-2^{-\frac{1}{3}\exp(n(\epsilon-\delta_n))},
\end{equation}
for all types $T_{X}\in\mathcal{P}_n(\mathcal{X})$ such that $R\leq D(T_{X}\|P_{X})+4\epsilon$. 
\end{enumerate}

\end{lemma}
\begin{IEEEproof}
We have
\begin{equation}
\mathbb{P}\left[\mathcal{B}_{1}(\epsilon|T_{X},P_{X})^{c}\right]=\mathbb{P}\left\{ \left|\frac{\varphi_{\mathcal{C}_{n}}(T_{X})}{\mathbb{E}\left[\varphi_{\mathcal{C}_{n}}(T_{X})\right]}-1\right|>2^{-n\epsilon}\right\} .
\end{equation}
Define $\theta_{m}(T_{X}):=\boldsymbol{1}{\{X^{n}(m)\in\mathcal{T}_{T_{X}}\}}$,
$m\in[2^{nR}]$ which are i.i.d. random variables, with mean 
\begin{equation}
p_{T_{X}}:=\mathbb{E}_{\mathcal{C}_{n}}\left[\theta_{m}(T_{X})\right]=P_X^{\times n}(\mathcal{T}_{T_X})\geq 2^{-n(D(T_{X}\|P_{X})+\delta_n)}.
\end{equation}
By identifying that $k  =2^{nR}$, $p =p_{T_{X}}$, $\delta =2^{-n\epsilon}$
and applying Lemma \ref{lem:chernoff-bound}, we obtain 
\begin{equation}
\mathbb{P}\left\{ \left|\frac{\varphi_{\mathcal{C}_{n}}(T_{X})}{\mathbb{E}\left[\varphi_{\mathcal{C}_{n}}(T_{X})\right]}-1\right|>2^{-n\epsilon}\right\} \leq 2^{-\frac{1}{3}\exp(n(\gamma-\delta_n))},\label{eq:double-fast}
\end{equation}
where $\gamma=R-D(T_{X}\|P_{X})-2\epsilon$. For fixed $\epsilon$
and $R\geq D(T_{X}\|P_{X})+4\epsilon$, it holds that $\gamma\geq\epsilon$.
Hence Eq.~\eqref{eq:double-fast} vanishes doubly exponentially
fast. This completes the proof of Statement 1.

Statement 2 follows from Statement 1 directly, since
\begin{align}
\mathbb{P}\left[\mathcal{B}_{2}(\epsilon|T_{X},P_{X})\right]=&\mathbb{P}\left[ \varphi_{\mathcal{C}_{n}}(T_{X})\leq 2\cdot2^{4n\epsilon}\right] \nonumber\\
\geq& \mathbb{P}\left\{\varphi_{\mathcal{C}_{n}}(T_{X})\leq (1+2^{-n\epsilon})\mathbb{E}\left[\varphi_{\mathcal{C}_{n}}(T_{X})\right] \right\}\nonumber\\
\geq& \mathbb{P}\left[\mathcal{B}_{1}(\epsilon|T_{X},P_{X})\right],
\end{align}
where in the first inequality we have used Eq.~\eqref{eq:size-mean-1}.
\end{IEEEproof}

For any type $T_{X}\in\mathcal{P}_{n}(\mathcal{X})$ and probability
distribution $P_{X}$ such that $R-D(T_{X}\|P_{X})\geq4\epsilon$,
define two events on $\mathcal{C}_{n}$ as 
\begin{align}
\mathcal{B}_{3}(\epsilon|T_{X},P_{X}) & :=\bigcap_{T_{Y|X}:D(T_{X}\|P_{X})+I(X:Y)_{T_{XY}}\leq R-4\epsilon}\mathcal{B}_{3}(\epsilon|T_{XY},P_{X}),\nonumber \\
\mathcal{B}_{4}(\epsilon|T_{X},P_{X}) & :=\bigcap_{T_{Y|X}:D(T_{X}\|P_{X})+I(X:Y)_{T_{XY}}\geq R-4\epsilon}\mathcal{B}_{4}(\epsilon|T_{XY},P_{X}).
\end{align}

\begin{lemma}[Strong Packing-Covering Lemma for Constant Composition
Codes~\cite{Yu2024renyi}] \label{lem:packing-coveing}
Let $\epsilon>0$. It holds that 
\begin{equation}
\mathbb{P}\left[\mathcal{B}_{3}(\epsilon|T_{X},P_{X})\cap\mathcal{B}_{4}(\epsilon|T_{X},P_{X})\big|\mathcal{B}_{1}(\epsilon|T_{X},P_{X})\right]\geq1-2^{-\exp(n(\epsilon-o_{n}(1))}
\end{equation}
for all pairs $(T_{X},P_{X})\in\mathcal{P}_{n}(\mathcal{X})\times\mathcal{P(X)}$
such that $R-D(T_{X}\|P_{X})\geq4\epsilon$, where $o_{n}(1)$ is
a term independent of $(T_{X},P_{X},R)$ and vanishes as $n\to\infty$.
That is, the probability above converges to one doubly exponentially
fast for all pairs $(T_{X},P_{X})\in\mathcal{P}_{n}(\mathcal{X})\times\mathcal{P(X)}$
such that $R-D(T_{X}\|P_{X})\geq4\epsilon$ as $n\to\infty$. 
\end{lemma}

\begin{remark}
Rigorously speaking, the original
version of \cite[Lemma 8]{Yu2024renyi} considers constant composition
codes, i.e., a codebook consisting of $2^{nR'}$ i.i.d. codewords
$X^{n}\sim\mathrm{Unif}(\mathcal{T}_{T_{X}})$, but under the condition
that $\mathcal{B}_{1}(\epsilon|T_{X},P_{X})$ occurs, the number of
codewords in our codebook $\mathcal{C}_{n}$ falling in $\mathcal{T}_{T_{X}}$,
i.e., $\varphi_{\mathcal{C}_{n}}(T_{X})$, is sandwiched between $2^{n(R-D(T_{X}\|P_{X})-\epsilon+o_{n}(1))}$
and $2^{n(R-D(T_{X}\|P_{X})+\epsilon+o_{n}(1))}$, not exactly equal
to $2^{n(R-D(T_{X}\|P_{X}))}$. However, this subtle difference is
not important, since by checking the proof of \cite[Lemma 8]{Yu2024renyi}
(specifically by invoking \cite[Lemma 6]{Yu2024renyi}), the lemma
with slightly modification as done in the lemma above still works
in our setting.
\end{remark}

For $i=\{1,3,4\},$ define three events on $\mathcal{C}_{n}$ as
\begin{equation}
\mathcal{B}_{i}(\epsilon|P_{X}):=\bigcap_{T_{X}:R-D(T_{X}\|P_{X})\geq4\epsilon}\mathcal{B}_{i}(\epsilon|T_{X},P_{X}).
\end{equation}

\begin{lemma}[Strong Packing-Covering Lemma for I.I.D. Codes]\label{lem:packing-coveing-IID} 
Let $\epsilon>0$. It holds that
\begin{equation}
\mathbb{P}\left[\mathcal{B}_{1}(\epsilon|P_{X})\cap\mathcal{B}_{3}(\epsilon|P_{X})\cap\mathcal{B}_{4}(\epsilon|P_{X})\right]\geq1-2^{-\exp(n(\epsilon-o_{n}(1))}
\end{equation}
for all probability distribution $P_{X}$, where $o_{n}(1)$ is a
term independent of $(P_{X},R)$ and vanishes as $n\to\infty$. That
is, the probability above converges to one doubly exponentially fast
and uniformly for all probability distribution $P_{X}$ as $n\to\infty$. 
\end{lemma}
\begin{IEEEproof}
Using a union bound, we have
\begin{align}
& \mathbb{P}\left[\left(\mathcal{B}_{1}(\epsilon|P_{X})\cap\mathcal{B}_{3}(\epsilon|P_{X})\cap\mathcal{B}_{4}(\epsilon|P_{X})\right)^{c}\right]\nonumber \\
\leq & \mathbb{P}\left[\mathcal{B}_{1}(\epsilon|P_{X})^{c}\right]+\mathbb{P}\left[\left(\mathcal{B}_{3}(\epsilon|P_{X})\cap\mathcal{B}_{4}(\epsilon|P_{X})\right)^{c}|\mathcal{B}_{1}(\epsilon|P_{X})\right]\nonumber \\
\leq & \sum_{T_{X}\in\mathcal{P}_{n}(\mathcal{X})}\left(\mathbb{P}\left[\mathcal{B}_{1}(\epsilon|T_{X},P_{X})^{c}\right]+\mathbb{P}\left[\left(\mathcal{B}_{3}(\epsilon|T_{X},P_{X})\cap\mathcal{B}_{4}(\epsilon|T_{X},P_{X})\right)^{c}|\mathcal{B}_{1}(\epsilon|T_{X},P_{X})\right]\right)\nonumber \\
\to & 0\text{ doubly exponentially fast,}
\end{align}
where the last line follows since the number of types is polynomial
in $n$. 
\end{IEEEproof}

\subsection{Proof of the Achievability Part}
In this subsection, we prove the achievability part of Theorem \ref{thm:SC-strong-converse-exponent}.
\begin{proposition}\label{thm:SC-SCE-A}
Let $R\geq0$, $P_{XY}\in\mathcal{P}(\mathcal{X}\times\mathcal{Y})$ be a distribution, and $\mathcal{C}_{n}=\{X^{n}(m)\}_{m=1}^{2^{nR}}$ be an i.i.d. random code, where each codeword $X^{n}(m)$ is drawn independently according to $P_{X}^{\times n}$. It holds that
\begin{equation}
\limsup_{n\to\infty} \frac{1}{n}\mathsf{D}_{\beta}(P_{XY}^{\times n},\mathcal{C}_{n})\leq
\begin{cases}
\max\limits_{\beta\leq\alpha\leq1}\frac{\beta(1-\alpha)}{\alpha(1-\beta)}\left\{ \widetilde{I}_{\alpha,\beta}(X:Y)_{P_{XY}}-R\right\} , & \beta\in(0,1)\\
\left|I_{\beta}(X:Y)_{P_{XY}}-R\right|^{+}, & \beta\in[1,\infty).
\end{cases}
\end{equation}
\end{proposition}
\begin{IEEEproof}
Let $R\geq4\epsilon>0$. Denote $M=2^{nR}$, $\beta=1+s$ and $\mathcal{C}_n(T_X)=\mathcal{C}_n\cap\mathcal{T}_{T_{X}}$ for any $T_{X}\in\mathcal{P}_{n}(\mathcal{X})$. For each $m\in[M]$, set $f_{\mathcal{C}_n}(m)=X^n(m)$. Then for $s\in(-1,0)\cup(0,\infty)$, 
\begin{align}
& \exp\{s\mathsf{D}_{1+s}(P_{XY}^{\times n},\mathcal{C}_{n})\}\nonumber \\
= & \mathbb{E}_{\mathcal{C}_{n}}\sum_{y^{n}}\Big(\sum_{m}\frac{1}{M}P_{Y|X}^{\times n}(y^{n}|f_{\mathcal{C}_{n}}(m))\Big)^{1+s}(P_{Y}^{\times n}(y^{n}))^{-s}\nonumber \\
= & \mathbb{E}_{\mathcal{C}_{n}}\sum_{y^{n}}\Big(\sum_{T_{X|Y}}\sum_{m:f_{\mathcal{C}_{n}}(m)\in\mathcal{T}_{T_{X|Y}}(y^{n})}\frac{1}{M}P_{Y|X}^{\times n}(y^{n}|f_{\mathcal{C}_{n}}(m))\Big)^{1+s}(P_{Y}^{\times n}(y^{n}))^{-s}\nonumber \\
= & \mathbb{E}_{\mathcal{C}_{n}}\sum_{T_{Y}}\sum_{y^{n}\in\mathcal{T}_{T_{Y}}}2^{-(1+s)nR-sn\sum T_{Y}\!\log\!P_{Y}}\Big(\sum_{T_{X|Y}}2^{n\sum T_{XY}\!\log\!P_{Y|X}}\cdot\phi_{\mathcal{C}_{n}(T_{X})}(y^{n})\Big)^{1+s}\nonumber \\
\doteq & \mathbb{E}_{\mathcal{C}_{n}}\max_{T_{Y}}\sum_{y^{n}\in\mathcal{T}_{T_{Y}}}\max_{T_{X|Y}}2^{-sn\sum T_{Y}\!\log\!P_{Y}+(1+s)n(\sum T_{XY}\!\log\!P_{Y|X}-R)}\cdot\phi_{\mathcal{C}_{n}(T_{X})}^{1+s}(y^{n})\label{eq:-48}\\
\doteq & \mathbb{E}_{\mathcal{C}_{n}}\max_{T_{Y}}\sum_{y^{n}\in\mathcal{T}_{T_{Y}}}\sum_{T_{X|Y}}2^{-sn\sum T_{Y}\!\log\!P_{Y}+(1+s)n(\sum T_{XY}\!\log\!P_{Y|X}-R)}\cdot\phi_{\mathcal{C}_{n}(T_{X})}^{1+s}(y^{n})\label{eq:-1}\\
= & \mathbb{E}_{\mathcal{C}_{n}}\max_{T_{Y}}\sum_{T_{X|Y}}\sum_{y^{n}\in\mathcal{T}_{T_{Y}}}2^{-sn\sum T_{Y}\!\log\!P_{Y}+(1+s)n(\sum T_{XY}\!\log\!P_{Y|X}-R)}\cdot\phi_{\mathcal{C}_{n}(T_{X})}^{1+s}(y^{n})\nonumber \\
\doteq & \mathbb{E}_{\mathcal{C}_{n}}\max_{T_{XY}}\sum_{y^{n}\in\mathcal{T}_{T_{Y}}}2^{-sn\sum T_{Y}\!\log\!P_{Y}+(1+s)n(\sum T_{XY}\!\log\!P_{Y|X}-R)}\cdot\phi_{\mathcal{C}_{n}(T_{X})}^{1+s}(y^{n}),\label{eq:-51}
\end{align}
where $\phi_{\mathcal{C}_{n}(T_{X})}(y^{n})$ is defined in Eq.~\eqref{eq:definition-phi},
and Eqs.~\eqref{eq:-48}, \eqref{eq:-1} and \eqref{eq:-51} follow
since the numbers of types and conditional types are polynomial in
$n$ (cf.~Eqs.\eqref{eq:size-type} and \eqref{eq:size-conditional-type}).
Here, we use the shorthands $\sum T_{Y}\log T_{Y}\equiv\sum_{y}T_{Y}(y)\log T_{Y}(y)$
and $\sum T_{XY}\log P_{Y|X}\equiv\sum_{x,y}T_{XY}(x,y)\log P_{Y|X}(y|x)$.
In order to further estimate Eq.~\eqref{eq:-51}, we partition the set of types $T_{X}$ into two parts:
\begin{align}
\mathcal{T}_{1}:=&\{T_{X}:R<D(T_X\|P_X)+4\epsilon\}, \\
\mathcal{T}_{2}:=&\{T_{X}:R\geq D(T_X\|P_X)+4\epsilon\}.
\end{align}
So, the expression in Eq.~\eqref{eq:-51} lies between the minimum and maximum of $\eta_{1}$ and $\eta_{2}$, which are defined below.
\begin{equation}\label{eq:eta-1}
\eta_{1}:=\mathbb{E}_{\mathcal{C}_n} \max_{T_{XY}:T_{X}\in\mathcal{T}_{1}}\sum_{y^{n}\in\mathcal{T}_{T_{Y}}}2^{-sn\sum T_{Y}\!\log\! P_{Y}+(1+s)n(\sum T_{XY}\!\log\! P_{Y|X}-R)}\cdot\phi_{\mathcal{C}_n(T_X)}^{1+s}(y^n)
\end{equation}
and 
\begin{equation}
\eta_{2}:=\mathbb{E}_{\mathcal{C}_n} \max_{T_{XY}:T_{X}\in\mathcal{T}_{2}}\sum_{y^{n}\in\mathcal{T}_{T_{Y}}}2^{-sn\sum T_{Y}\!\log\! P_{Y}+(1+s)n(\sum T_{XY}\!\log\! P_{Y|X}-R)}\cdot\phi_{\mathcal{C}_n(T_X)}^{1+s}(y^n).
\end{equation}
The rest of the proof is divided into three cases.

\textit{Case 1:} $\beta\in(0,1)$. First, we estimate $\eta_{1}$. The key to estimating $\eta_{1}$ is to estimate the quantity
\begin{equation}
\mathbb{E}_{\mathcal{C}_n}\sum_{y^{n}\in\mathcal{T}_{T_{Y}}}\phi_{\mathcal{C}_n(T_X)}^{1+s}(y^n).
\end{equation}
Let $c$ be a realization of $\mathcal{C}_n$ such that $1\leq\varphi_{c}(T_X)\leq 2\cdot 2^{4n\epsilon}$. Denote $c(T_{X})=c\cap\mathcal{T}_{T_{X}}$ for any $T_{X}\in\mathcal{P}_{n}(\mathcal{X})$. It holds that
\begin{align}
\sum_{y^{n}\in\mathcal{T}_{T_{Y}}}\phi_{c(T_X)}^{1+s}(y^n)\geq& \sum_{y^{n}\in\mathcal{T}_{T_{Y}}} \boldsymbol{1}\Big\{y^{n}\in\bigcup_{x^{n}\in c(T_{X})}\mathcal{T}_{T_{Y|X}}(x^{n})\Big\} \nonumber\\
\geq&(n+1)^{-|\mathcal{X}|\cdot|\mathcal{Y}|} \cdot2^{nH(Y|X)_{T_{XY}}},
\end{align}
where the last inequality follows since 
\begin{equation}
\Big|\bigcup_{x^{n}\in c(T_{X})}\mathcal{T}_{T_{Y|X}}(x^{n})\Big|\geq (n+1)^{-|\mathcal{X}|\cdot|\mathcal{Y}|} \cdot2^{nH(Y|X)_{T_{XY}}}.
\end{equation}
Using Lemma~\ref{lem:iid}, we obtain that $\mathbb{P}\{\varphi_{\mathcal{C}_n}(T_X)> 2\cdot 2^{4n\epsilon}\}$ is doubly exponentially close to 0, where $\varphi_{\mathcal{C}_n}(T_X)$ is defined in Eq.~\eqref{eq:definition-varphi}. Together with Eq.~\eqref{eq:for-next-sc-1}, this yields
\begin{equation}
\mathbb{P}\{1\leq\varphi_{\mathcal{C}_n}(T_X)\leq 2\cdot 2^{4n\epsilon}\} \dot{\geq} 2^{n(R - D(T_X \| P_X)-4\epsilon)}.
\end{equation}
Hence, we obtain
\begin{align}\label{eq:key-eta-1-1}
&\mathbb{E}_{\mathcal{C}_n}\sum_{y^{n}\in\mathcal{T}_{T_{Y}}}\phi_{\mathcal{C}_n(T_X)}^{1+s}(y^n)\nonumber\\
\geq& \mathbb{P}\{1\leq\varphi_{\mathcal{C}_n}(T_X)\leq  2\cdot 2^{4n\epsilon}\}(n+1)^{-|\mathcal{X}|\cdot|\mathcal{Y}|} \cdot2^{nH(Y|X)_{T_{XY}}}\nonumber\\
\dot{\geq}&2^{nH(Y|X)_{T_{XY}}+n(R - D(T_X \| P_X)-4\epsilon)}
\end{align}
Then $\eta_1$ is lower bounded by
\begin{align}%\substack{T_{X}\in\mathcal{T}_2,\\T_{Y|X}\in\mathcal{T}_4}
\eta_{1}\dot{\geq}&\max_{T_{XY}:T_{X}\in\mathcal{T}_{1}}2^{-sn\sum T_{Y}\!\log\! P_{Y}+(1+s)(\sum T_{XY}\!\log\! P_{Y|X}-R)}\times 2^{nH(Y|X)_{T_{XY}}+n(R - D(T_X \| P_X)-4\epsilon)} \nonumber\\
=&\max_{T_{XY}:T_{X}\in\mathcal{T}_{1}}\!2^{-sn\sum T_{Y}\!\log\! P_{Y}+(1+s)n\sum T_{XY}\!\log\! P_{Y|X}+nH(Y|X)_{T_{XY}}}\times2^{-nD(T_X\|P_X)-snR-4n\epsilon} \nonumber\\
=&\max_{T_{XY}:T_{X}\in\mathcal{T}_{1}}2^{-(1+s)nD(T_{Y|X}\|P_{Y|X}|T_{X})+snD(T_{Y|X}\|P_{Y}|T_{X})}\times2^{-nD(T_{X}\|P_{X})-snR-4n\epsilon}\nonumber \\
=&\max_{T_{XY}:T_{X}\in\mathcal{T}_{1}}2^{-(1+s)nD(T_{XY}\|P_{XY})+snD(T_{Y|X}\|P_{Y}|T_{X})+sn(D(T_{X}\|P_{X})-R)-4n\epsilon}\nonumber\\
=&\max_{T_{XY}:T_{X}\in\mathcal{T}_{1}}2^{snD(T_{Y}\|P_{Y})-(1+s)n(D(T_{XY}\|P_{XY})+4\epsilon)}\times2^{sn(D(T_{X|Y}\|P_{X}|T_Y)-R+4\epsilon)}, \label{eq:sc-final-1}
\end{align}

Next, we estimate $\eta_{2}$. By Lemma \ref{lem:packing-coveing-IID},
there is a realization $c'$ of $\mathcal{C}_{n}$ satisfying $\mathcal{B}_{1}(\epsilon|P_{X})\cap\mathcal{B}_{2}(\epsilon|P_{X})\cap\mathcal{B}_{3}(\epsilon|P_{X})$.
In fact, this happens with probability doubly exponentially close to
$1$. Let $R_{T_{X}}$ be such that $|c'(T_{X})|=2^{nR_{T_{X}}}$. From the definition of $\mathcal{B}_{1}(\epsilon|P_{X})$, we obtain
\begin{equation}\label{eq:sc-RTX}
R-D(T_X\|P_X) - \frac{1}{n}\log(|\mathcal{X}|\cdot|\mathcal{Y}|)\leq R_{T_{X}}\leq R-D(T_X\|P_X) +\frac{1}{n}.
\end{equation}
In order to estimate $\eta_2$,
we partition the set of conditional types $T_{Y|X}$ into two parts:
\begin{align}
\mathcal{T}_{3}:=&\{T_{Y|X}:I(X:Y)_{T_{XY}}\le R_{T_{X}}-4\epsilon\},\\
\mathcal{T}_{4}:=&\{T_{Y|X}:I(X:Y)_{T_{XY}}\ge R_{T_{X}}-4\epsilon\}.
\end{align}
By Lemma \ref{lem:packing-coveing-IID}, for all $T_{Y|X}\in\mathcal{T}_{3}$
and $y^{n}\in\mathcal{T}_{T_{Y}}$, 
\begin{equation}
(n+1)^{-|\mathcal{X}|\cdot|\mathcal{Y}|}2^{n\left(R_{T_X}-I(X:Y)_{T_{XY}}-4\epsilon\right)}\leq \phi_{c'(T_X)}(y^n)\leq 2^{n(R_{T_{X}}-I(X:Y)_{T_{XY}}+4\epsilon)};
\end{equation}
and for all $T_{Y|X}\in\mathcal{T}_{4}$ and $y^{n}\in\mathcal{T}_{T_{Y}}$,
\begin{equation}
\boldsymbol{1}\Big\{y^{n}\in\bigcup_{x^{n}\in c'(T_{X})}\mathcal{T}_{T_{Y|X}}(x^{n})\Big\}\leq\phi_{c'(T_X)}(y^n)\leq 2^{5n\epsilon}\cdot\boldsymbol{1}\Big\{y^{n}\in\bigcup_{x^{n}\in c'(T_{X})}\mathcal{T}_{T_{Y|X}}(x^{n})\Big\}.
\end{equation}
Define
\begin{equation}\label{eq:gamma}
\gamma:=\max_{T_{XY}:T_{X}\in\mathcal{T}_{2}}\sum_{y^{n}\in\mathcal{T}_{T_{Y}}}2^{-sn\sum T_{Y}\!\log\! P_{Y}+(1+s)n(\sum T_{XY}\!\log\! P_{Y|X}-R)}\cdot\phi_{c'(T_X)}^{1+s}(y^n).
\end{equation}
So, $\gamma$ is lower bounded by the minimum
of $\gamma_{1}$ and $\gamma_{2}$ defined below.
\begin{align}%\max_{\substack{T_{X}\in\mathcal{T}_2,\\T_{Y|X}\in\mathcal{T}_3}}
\gamma_{1}  :=&\max_{T_{X}\in\mathcal{T}_2,T_{Y|X}\in\mathcal{T}_3}\sum_{y^{n}\in\mathcal{T}_{T_{Y}}}\!\!2^{-sn\sum T_{Y}\!\log\! P_{Y}+(1+s)n(\sum T_{XY}\!\log\! P_{Y|X}-R)}\cdot(n+1)^{-(1+s)|\mathcal{X}|\cdot|\mathcal{Y}|}2^{(1+s)n\left(R_{T_X}-I(X:Y)_{T_{XY}}-4\epsilon\right)} \nonumber\\
\dot{\geq}&\max_{T_{X}\in\mathcal{T}_2,T_{Y|X}\in\mathcal{T}_3} 2^{nH(Y)_{T_{Y}}-sn\sum T_{Y}\!\log\! P_{Y}+(1+s)n(\sum T_{XY}\!\log\! P_{Y|X}-D(T_{X}\|P_{X})-I(X:Y)_{T_{XY}}-4\epsilon)}\label{eq:gamma-1-1}\\
=&\max_{T_{X}\in\mathcal{T}_2,T_{Y|X}\in\mathcal{T}_3} 2^{snD(T_{Y}\|P_{Y})-(1+s)n(D(T_{XY}\|P_{XY})+4\epsilon)},\label{eq:sc-gamma1-1}
\end{align}
and 
\begin{align}
\gamma_{2}:= & \max_{T_{X}\in\mathcal{T}_2,T_{Y|X}\in\mathcal{T}_4} \sum_{y^{n}\in\mathcal{T}_{T_{Y}}}2^{-sn\sum T_{Y}\!\log\! P_{Y}+(1+s)n(\sum T_{XY}\!\log\! P_{Y|X}-R)}\cdot\boldsymbol{1}\Big\{y^{n}\in\bigcup_{x^{n}\in c'(T_{X})}\mathcal{T}_{T_{Y|X}}(x^{n})\Big\} \nonumber\\
\dot{\geq} & \max_{T_{X}\in\mathcal{T}_2,T_{Y|X}\in\mathcal{T}_4} 2^{-sn\sum T_{Y}\!\log\! P_{Y}+(1+s)n(\sum T_{XY}\!\log\! P_{Y|X}-R)+n(R_{T_{X}}+H(Y|X)_{T_{XY}})}\label{eq:-2-1-1}\\
\doteq & \max_{T_{X}\in\mathcal{T}_2,T_{Y|X}\in\mathcal{T}_4} 2^{-(1+s)nD(T_{Y|X}\|P_{Y|X}|T_{X})+snD(T_{Y|X}\|P_{Y}|T_{X})-nD(T_{X}\|P_{X})-snR} \label{eq:sc-gamma2-2}\\
= & \max_{T_{X}\in\mathcal{T}_2,T_{Y|X}\in\mathcal{T}_4} 2^{-(1+s)nD(T_{XY}\|P_{XY})+snD(T_{Y|X}\|P_{Y}|T_{X})+sn(D(T_{X}\|P_{X})-R)} \nonumber\\
= & \max_{T_{X}\in\mathcal{T}_2,T_{Y|X}\in\mathcal{T}_4} 2^{snD(T_{Y}\|P_{Y})-(1+s)nD(T_{XY}\|P_{XY})+sn(D(T_{X|Y}\|P_{X}|T_Y)-R)}, \label{eq:sc-gamma2}
\end{align}
where Eq.~\eqref{eq:gamma-1-1} follows from Eqs.~\eqref{eq:type-size} and \eqref{eq:sc-RTX}, Eq.~\eqref{eq:-2-1-1} is derived by the inequality
\begin{equation}
\Big|\bigcup_{x^{n}\in c'(T_{X})}\mathcal{T}_{T_{Y|X}}(x^{n})\Big|\geq (n+1)^{-|\mathcal{X}|\cdot|\mathcal{Y}|} \cdot2^{n(R_{T_X}+H(Y|X)_{T_{XY}})},
\end{equation}
and Eq.~\eqref{eq:sc-gamma2-2} comes from Eq.~\eqref{eq:sc-RTX}.
Combining Eqs.~\eqref{eq:sc-gamma1-1} and \eqref{eq:sc-gamma2}, we have
\begin{align}
\gamma\dot{\geq}&\max_{T_{XY}:T_{X}\in\mathcal{T}_{2}}2^{snD(T_{Y}\|P_{Y})-(1+s)n(D(T_{XY}\|P_{XY})+4\epsilon)}\min\{1,2^{sn(D(T_{X|Y}\|P_{X}|T_Y)-R)}\} \nonumber\\
\geq&\max_{T_{XY}:T_{X}\in\mathcal{T}_{2}}2^{snD(T_{Y}\|P_{Y})-(1+s)n(D(T_{XY}\|P_{XY})+4\epsilon)}\min\{1,2^{sn(D(T_{X|Y}\|P_{X}|T_Y)-R+4\epsilon)}\}.
\end{align} 
Lemma \ref{lem:packing-coveing-IID} shows that with probability doubly exponentially close to $1$, $\mathcal{C}_{n}$
takes such realizations $c'$. Thus, it holds that
\begin{equation}\label{eq:sc-final-2}
\eta_{2}\dot{\geq}\max_{T_{XY}:T_{X}\in\mathcal{T}_{2}}2^{snD(T_{Y}\|P_{Y})-(1+s)n(D(T_{XY}\|P_{XY})+4\epsilon)}\min\{1,2^{sn(D(T_{X|Y}\|P_{X}|T_Y)-R+4\epsilon)}\}.
\end{equation}

Since $\mathcal{P}_n(\mathcal{X}\times \mathcal{Y})$ is dense in $\mathcal{P}(\mathcal{X}\times \mathcal{Y})$ as $n \to \infty$, the combination of Eqs.~\eqref{eq:sc-final-1} and \eqref{eq:sc-final-2} yields
\begin{align}
&\limsup_{n \to \infty}\frac{1}{n}\mathsf{D}_{1+s}(P_{XY}^{\times n},\mathcal{C}_n) \nonumber\\
\leq&
\min_{T_{XY}\in\mathcal{P}(\mathcal{X}\times \mathcal{Y})}\Big\{D(T_{Y}\|P_{Y})-\frac{1+s}{s}D(T_{XY}\|P_{XY})+|D(T_{X|Y}\|P_{X}|T_Y)-R+4\epsilon|^{+}\Big\}-4\epsilon\frac{1+s}{s}\nonumber \\
\leq&\min_{T_{XY}\in\mathcal{P}(\mathcal{X}\times \mathcal{Y})}\Big\{D(T_{Y}\|P_{Y})-\frac{1+s}{s}D(T_{XY}\|P_{XY}) +|D(T_{X|Y}\|P_{X}|T_Y)-R|^{+}\Big\}-\frac{4\epsilon}{s} \nonumber\\
=&\max_{\alpha\in[\beta,1]}\frac{\beta(1-\alpha)}{\alpha(1-\beta)}\Big\{\widetilde{I}_{\alpha,\beta}(X:Y)_{P_{XY}}-R\Big\}-\frac{4\epsilon}{s},
\end{align}
where in the last equality we apply Lemma~\ref{lem:vari-sc} (which
will be given later). The desired result then follows by letting $\epsilon\searrow0$.

\textit{Case 2:} $\beta\in(1,\infty)$. By an argument similar to the case $\beta\in(0,1)$, the quantities $\eta_{1}$ and $\eta_{2}$ can be bounded as
\begin{align}
\eta_{1}\dot{\leq}&\max_{T_{XY}:T_{X}\in\mathcal{T}_{1}}2^{snD(T_{Y}\|P_{Y})-(1+s)nD(T_{XY}\|P_{XY})+8n\epsilon}\cdot 2^{sn(D(T_{X|Y}\|P_{X}|T_Y)-R+4\epsilon)}, \label{eq:sc-final-1-1}\\ 
\eta_{2}\dot{\leq}&\max_{T_{XY}:T_{X}\in\mathcal{T}_{2}}2^{snD(T_{Y}\|P_{Y})-(1+s)n(D(T_{XY}\|P_{XY})-5\epsilon)}\max\{1,2^{sn(D(T_{X|Y}\|P_{X}|T_Y)-R+4\epsilon)}\}.\label{eq:sc-final-2-1}
\end{align}
A detailed proof of these bounds is provided in Appendix~\ref{app:eta-1-2}.
The combination of Eqs.~\eqref{eq:sc-final-1-1} and \eqref{eq:sc-final-2-1} yields
\begin{align}
&\limsup_{n \to \infty}\frac{1}{n}\mathsf{D}_{1+s}(P_{XY}^{\times n},\mathcal{C}_n) \nonumber\\
\leq&\max _{T_{XY}\in\mathcal{P}_n(\mathcal{X}\times \mathcal{Y})}\Big\{D(T_{Y}\|P_{Y})-\frac{1\!+\!s}{s}D(T_{XY}\|P_{XY})+|D(T_{X|Y}\|P_{X}|T_Y)-R+4\epsilon|^{+}\Big\}\!+\!5\epsilon\frac{2\!+\!s}{s}\nonumber \\
\leq&\max_{T_{XY}\in\mathcal{P}(\mathcal{X}\times \mathcal{Y})}\Big\{D(T_{Y}\|P_{Y})- \frac{1\!+\!s}{s}D(T_{XY}\|P_{XY})+|D(T_{X|Y}\|P_{X}|T_Y)-R|^{+}\Big\}\!+\!10\epsilon\frac{1\!+\!s}{s} \nonumber\\
\leq&\max_{T_{XY}}\max\Big\{\frac{1\!+\!s}{-s}D(T_{XY}\|P_{XY})+D(T_{XY}\|P_{X}\times P_{Y})-R,\frac{1\!+\!s}{-s}D(T_{XY}\|P_{XY})\!+\!D(T_{Y}\|P_{Y})\Big\} \!+\!10\epsilon\frac{1\!+\!s}{s} \nonumber\\
=&\max\Big\{\max_{T_{XY}}\Big\{\frac{1\!+\!s}{-s}D(T_{XY}\|P_{XY})\!+\!D(T_{XY}\|P_{X}\times P_{Y})\!-\!R\Big\}, \nonumber\\
&\qquad\qquad\qquad\qquad\qquad\max_{T_{XY}}\Big\{\frac{1\!+\!s}{-s}D(T_{XY}\|P_{XY})\!+\!D(T_{Y}\|P_{Y})\Big\}\Big\} \!+\!10\epsilon\frac{1\!+\!s}{s} \nonumber\\
=&\max\{D_\beta(P_{XY}\|P_X\times P_Y)-R,0\} +10\epsilon\frac{1+s}{s} \nonumber\\
=&\left|I_{\beta}(X:Y)_{P_{XY}}-R\right|^{+}+10\epsilon\frac{1+s}{s},
\end{align}
where the second equality follows from the variational expression of R{\'e}nyi divergence (Lemma~\ref{lem:RenyiD-properties}). Letting $\epsilon \searrow 0$ yields the desired result.

\textit{Case 3:} $\beta=1$. We will show that the desired result follows from the results of case $\beta\in(1,\infty)$. By the monotonicity of R{\'e}nyi divergence (Lemma~\ref{lem:RenyiD-properties}), for any $\beta>1$ we have
\begin{align}\label{eq:beta-eq-1}
\limsup_{n \to \infty}\frac{1}{n}\mathsf{D}(P_{XY}^{\times n},\mathcal{C}_n)\leq& \limsup_{n \to \infty}\frac{1}{n}\mathsf{D}_{\beta}(P_{XY}^{\times n},\mathcal{C}_n)\nonumber\\
\leq&\left|I_{\beta}(X:Y)_{P_{XY}}-R\right|^{+}.
\end{align}
Using the continuity of the function $\beta\mapsto|I_\beta(X:Y)_{P_{XY}}-R|^+$ and $\lim_{\beta\to1}I_\beta(X:Y)_{P_{XY}}=I(X:Y)_{P_{XY}}$, we get
\begin{equation}\label{eq:beta-eq-2}
\lim_{\beta \searrow 1}\limsup_{n \to \infty}\frac{1}{n}\mathsf{D}_{\beta}(P_{XY}^{\times n},\mathcal{C}_n)\leq|I(X:Y)_{P_{XY}}-R|^+.
\end{equation}
Combining Eqs.~\eqref{eq:beta-eq-1} and \eqref{eq:beta-eq-2}, the desired result follows. This completes the proof.
\end{IEEEproof}
In the proof of Proposition~\ref{thm:SC-SCE-A}, we have used the following variational expression. 
\begin{lemma}\label{lem:vari-sc}
Let $\beta\in(0,1)$, $R\geq0$ and $P_{XY}\in\mathcal{P}(\mathcal{X}\times\mathcal{Y})$.
It holds that
\begin{align}
&\max_{\alpha\in[\beta,1]}\frac{\beta(1-\alpha)}{\alpha(1-\beta)}\Big\{\widetilde{I}_{\alpha,\beta}(X:Y)_{P_{XY}}-R\Big\} \nonumber\\
=&\min_{Q_{XY}\in\mathcal{P}(\mathcal{X}\times\mathcal{Y})}\Big\{D(Q_{Y}\|P_{Y})+\frac{\beta}{1-\beta}D(Q_{XY}\|P_{XY})+\left|D(Q_{X|Y}\|P_{X}|Q_{Y})-R\right|^{+}\Big\}.
\end{align}
\end{lemma}
\begin{IEEEproof}
For any $\alpha\in[\beta,1)$, applying Theorem~\ref{thm:variational-expression-I} yields
\begin{align}\label{eq:re-two}
&\frac{\beta(1-\alpha)}{\alpha(1-\beta)}\Big\{\widetilde{I}_{\alpha,\beta}(X:Y)_{P_{XY}}-R\Big\}\nonumber \\
= & \min_{Q_{XY}\in\mathcal{P}(\mathcal{X}\times\mathcal{Y})}\frac{\beta(1-\alpha)}{\alpha(1-\beta)}\Big\{\frac{\alpha(1-\beta)}{\beta(1-\alpha)}D(Q_{Y}\|P_{Y})+\frac{\alpha}{1-\alpha}D(Q_{XY}\|P_{XY})+D(Q_{X|Y}\|P_{X}|Q_{Y})-R\Big\}\nonumber \\
=&\min_{Q_{XY}\in\mathcal{P}(\mathcal{X}\times\mathcal{Y})}\Big\{D(Q_{Y}\|P_{Y})+\frac{\beta}{1-\beta}D(Q_{XY}\|P_{XY})+\frac{\beta(1-\alpha)}{\alpha(1-\beta)}\left(D(Q_{X|Y}\|P_{X}|Q_{Y})-R\right)\Big\}.
\end{align}
When $\alpha = 1$, the first and last expressions in Eq.~\eqref{eq:re-two} coincide, as they are both $0$. Therefore, we get
\begin{align}
& \max_{\alpha\in[\beta,1]}\frac{\beta(1-\alpha)}{\alpha(1-\beta)}\Big\{\widetilde{I}_{\alpha,\beta}(X:Y)_{P_{XY}}-R\Big\}\nonumber \\
=& \max_{\alpha\in[\beta,1]}\min_{Q_{XY}\in\mathcal{P}(\mathcal{X}\times\mathcal{Y})}\Big\{D(Q_{Y}\|P_{Y})+\frac{\beta}{1-\beta}D(Q_{XY}\|P_{XY})+\frac{\beta(1-\alpha)}{\alpha(1-\beta)}\left(D(Q_{X|Y}\|P_{X}|Q_{Y})-R\right)\Big\}\nonumber \\
\overset{(a)}{=} & \max_{\lambda\in[0,1]}\min_{Q_{XY}\in\mathcal{P}(\mathcal{X}\times\mathcal{Y})}\Big\{D(Q_{Y}\|P_{Y})+\frac{\beta}{1-\beta}D(Q_{XY}\|P_{XY})+\lambda\left(D(Q_{X|Y}\|P_{X}|Q_{Y})-R\right)\Big\}\nonumber \\
\overset{(b)}{=} & \min_{Q_{XY}\in\mathcal{P}(\mathcal{X}\times\mathcal{Y})}\max_{\lambda\in[0,1]}\Big\{D(Q_{Y}\|P_{Y})+\frac{\beta}{1-\beta}D(Q_{XY}\|P_{XY})+\lambda\left(D(Q_{X|Y}\|P_{X}|Q_{Y})-R\right)\Big\}\nonumber \\
= & \min_{Q_{XY}\in\mathcal{P}(\mathcal{X}\times\mathcal{Y})}\Big\{D(Q_{Y}\|P_{Y})+\frac{\beta}{1-\beta}D(Q_{XY}\|P_{XY})+\left|D(Q_{X|Y}\|P_{X}|Q_{Y})-R\right|^{+}\Big\},
\end{align}
where $(a)$ is by setting $\frac{\beta(1-\alpha)}{\alpha(1-\beta)}=\lambda$
and $(b)$ comes from Sion's minimax theorem. To verify the applicability of Sion's minimax theorem here, note that (i) the function $\lambda\mapsto\lambda(D(Q_{X|Y}\|P_{X}|Q_{Y})-R)$
is linear and continuous, and (ii) the function $Q_{XY}\mapsto D(Q_{Y}\|P_{Y})+\frac{\beta}{1-\beta}D(Q_{XY}\|P_{XY})+\lambda(D(Q_{X|Y}\|P_{X}|Q_{Y})-R)$ is convex and lower semi-continuous. 
\end{IEEEproof}

\subsection{Proof of the Optimality Part}
In this subsection, we prove a one-shot version of the optimality part. Then, we apply it directly to deal with the asymptotic situation. 
\begin{proposition}\label{prop:one-shot-sc}
Let $\mathcal{C}=\{X(m)\}_{m=1}^{M}$ be a random code, where each codeword $X(m)$ is independently drawn from $P_{X}$. For any channel $P_{Y|X}$, we have
\begin{equation}
\mathsf{D}_{\beta}(P_{XY},\mathcal{C})\geq\begin{cases}
\max\limits_{\beta\leq\alpha\leq1}\frac{\beta(1-\alpha)}{\alpha(1-\beta)}\left\{ \widetilde{I}_{\alpha,\beta}(X:Y)_{P_{XY}}-\log M\right\} , & \beta\in(0,1)\\
\left|I_{\beta}(X:Y)_{P_{XY}}-\log M\right|^{+}, & \beta\geq1.
\end{cases}
\end{equation}
\end{proposition}
\begin{IEEEproof}
At first, we consider the case $\beta\in(0,1)$. Let $\alpha\in[\beta,1]$
and set $f_{\mathcal{C}}(m)=X(m)$. Lemma \ref{lem:-2} (given in
Appendix \ref{app:Miscellaneous-Lemmas}) implies that 
\begin{equation}
\Big(\sum_{m=1}^{M}P_{Y|X}^{\alpha}(y|f_{\mathcal{C}}(m))\Big)^{\frac{1}{\alpha}}\geq\sum_{m=1}^{M}P_{Y|X}(y|f_{\mathcal{C}}(m)).
\end{equation}
Using this relation and Jensen's inequality, we obtain that
\begin{align}
\mathbb{E}_{\mathcal{C}} P_{Y|\mathcal{C}}^\beta(y)
=& \mathbb{E}_{\mathcal{C}}\Big(\sum_{m=1}^{M}\frac{1}{M} P_{Y|X}(y|f_{\mathcal{C}}(m))\Big)^{\beta}\nonumber \\
\leq & M^{-\beta}\mathbb{E}_{\mathcal{C}}\Big(\sum_{m=1}^{M} P_{Y|X}^{\alpha}(y|f_{\mathcal{C}}(m))\Big)^{\frac{\beta}{\alpha}}\nonumber \\
\leq & M^{-\beta}\Big(\mathbb{E}_{\mathcal{C}}\sum_{m=1}^{M}P_{Y|X}^{\alpha}(y|f_{\mathcal{C}}(m))\Big)^{\frac{\beta}{\alpha}}\nonumber \\
= & M^{\frac{\left(1-\alpha\right)\beta}{\alpha}}\Big(\sum_{x}P_{X}(x)P_{Y|X}^{\alpha}(y|x)\Big)^{\frac{\beta}{\alpha}}.\label{eq:one-shot-sc}
\end{align}
Combining Eqs.\eqref{eq:definition-one-shot} and \eqref{eq:one-shot-sc}, we obtain
\begin{align*}
D_{\beta}(P_{XY},\mathcal{C})\!
\geq & \frac{1}{\beta\!-\!1}\log\sum_{y}P_{Y}(y)\Big(\!\sum_{x}\!P_{X}^{1-\alpha}(x)P_{X|Y}^{\alpha}(x|y)\!\Big)^{\frac{\beta}{\alpha}}-\frac{\beta(1-\alpha)}{\alpha(1-\beta)}\log M\nonumber \\
= & \frac{\beta(1-\alpha)}{\alpha(1-\beta)}\left\{\widetilde{I}_{\alpha,\beta}(X:Y)_{P_{XY}}-\log M\right\}.
\end{align*}
This leads to the statement for $\beta\in(0,1)$. For the case $\beta>1$, Lemma \ref{lem:-2} implies that 
\begin{align}
& D_{\beta}(P_{XY},\mathcal{C})\nonumber \\
= & \frac{1}{\beta-1}\log\sum_{y}P_{Y}^{1-\beta}(y)\mathbb{E}_{\mathcal{C}}\Big(\sum_{m=1}^{M}\frac{1}{M}P_{Y|X}(y|f_{\mathcal{C}}(m))\Big)^{\beta}\nonumber \\
\geq & \frac{1}{\beta-1}\log M^{-\beta}\sum_{y}P_{Y}^{1-\beta}(y)\sum_{m=1}^{M}\mathbb{E}_{\mathcal{C}}P_{Y|X}^{\beta}(y|f_{\mathcal{C}}(m))\nonumber \\
= & \frac{1}{\beta-1}\log \sum_{y}P_{Y}^{1-\beta}(y)\sum_{x}P_{X}(x)P_{Y|X}^{\beta}(y|x)-\log M\nonumber \\
= & I_{\beta}(X:Y)_{P_{XY}}-\log M.
\end{align}
Because the R{\'e}nyi divergence is non-negative, we get the desired result.
At last, taking the limit $\beta\to1$, we obtain the result for $\beta=1$. 
\end{IEEEproof}
Lemma~\ref{prop:one-shot-sc} directly implies the following corollary,
which completes the proof of the optimality part of Theorem~\ref{thm:SC-strong-converse-exponent}.
\begin{corollary}
Let $R\geq0$, $P_{XY}\in\mathcal{P}(\mathcal{X}\times\mathcal{Y})$ be a distribution, and let $\mathcal{C}_{n}=\{X^{n}(m)\}_{m=1}^{2^{nR}}$ be an i.i.d. random code, where each codeword $X^{n}(m)$ is drawn independently according to $P_{X}^{\times n}$. It holds that
\begin{equation}
\liminf_{n\to\infty} \frac{1}{n}\mathsf{D}_{\beta}(P_{XY}^{\times n},\mathcal{C}_{n})\geq \begin{cases}
\max\limits_{\beta\leq\alpha\leq1}\frac{\beta(1-\alpha)}{\alpha(1-\beta)}\left\{ \widetilde{I}_{\alpha,\beta}(X:Y)_{P_{XY}}-\log M\right\} , & \beta\in(0,1)\\
\left|I_{\beta}(X:Y)_{P_{XY}}-\log M\right|^{+}, & \beta\geq1.
\end{cases}
\end{equation}
\end{corollary}
\begin{IEEEproof}
Since $\widetilde{I}_{\alpha,\beta}(X:Y)_{P_{XY}}$ and $I_{\beta}(X:Y)_{P_{XY}}$
are additive (Proposition~\ref{prop:additivity-I}), by applying Lemma~\ref{prop:one-shot-sc} with the substitutions $\mathcal{C}\leftarrow\mathcal{C}_{n}$, $P_{XY}\leftarrow P_{XY}^{\times n}$, and $M\leftarrow 2^{nR}$, we obtain the desired result.
\end{IEEEproof}

\section{Conclusion and Discussion}\label{sec:con-dis}
In this paper, we study a two-parameter R{\'e}nyi
conditional entropy introduced in \cite{RGT2024quantum, HayashiTan2016equivocations}
and investigate its limiting behavior as the parameters approach zero
or infinity, showing that it recovers two existing definitions. We
further introduce a two-parameter R{\'e}nyi mutual information, which
unifies several existing definitions within a single framework. We
examine fundamental properties of these two-parameter quantities,
including monotonicity with respect to the R{\'e}nyi parameters, additivity,
data-processing inequalities, and variational expressions. Finally,
we apply these quantities to characterize the strong converse exponents
in privacy amplification and soft covering problems.

Regarding the limiting cases where the parameters approach zero, due
to that continuous extension fails in this case, it is unclear which are  the most
reasonable and consistent definitions in this case. 
%These
%boundary cases may yield distinct forms of conditional entropy and
%mutual information. 
Clarifying these limits could lead to a deeper
understanding of extreme regimes in information measures.

The other important direction is to explore whether the two-parameter R{\'e}nyi information quantities admit any operational interpretations in the parameter region  where the operation explanations are currently lacking.

\section*{Acknowledgements}
The authors would like to thank Zhiwen Lin for bringing to their attention Reference~\cite{RGT2024quantum}.

\appendices{}
\section{Upper Bound of $\eta_1$ and $\eta_2$ in Theorem~\ref{thm:SC-strong-converse-exponent}}
\label{app:eta-1-2} \setcounter{equation}{0} 
\global\long\def\theequation{A.\arabic{equation}}%
\begin{IEEEproof}
We first estimate $\eta_1$. The key step is to evaluate 
\begin{equation}
\mathbb{E}_{\mathcal{C}_n}\sum_{y^{n}\in\mathcal{T}_{T_{Y}}}\phi_{\mathcal{C}_n(T_X)}^{1+s}(y^n).
\end{equation}
Let $c$ be a realization of $\mathcal{C}_n$ such that $1\leq\varphi_{c}(T_X)\leq 2\cdot 2^{4n\epsilon}$. It holds that
\begin{align}
\sum_{y^{n}\in\mathcal{T}_{T_{Y}}}\phi_{c(T_X)}^{1+s}(y^n)\leq& \sum_{y^{n}\in\mathcal{T}_{T_{Y}}}  2^{1+s}\cdot  2^{4(1+s)n\epsilon}\cdot\boldsymbol{1}\Big\{y^{n}\in\bigcup_{x^{n}\in c(T_{X})}\mathcal{T}_{T_{Y|X}}(x^{n})\Big\} \nonumber\\
\leq&2^{1+s}\cdot 2^{(2+s)4n\epsilon+nH(Y|X)_{T_{XY}}},
\end{align}
where the last inequality follows since 
\begin{equation}
\Big|\bigcup_{x^{n}\in c(T_{X})}\mathcal{T}_{T_{Y|X}}(x^{n})\Big|\leq 2^{n(4\epsilon+H(Y|X)_{T_{XY}})}.
\end{equation}
From Eq.~\eqref{eq:for-next-sc}, we have
\begin{equation}
\mathbb{P}\{1\leq\varphi_{\mathcal{C}_n}(T_X)\leq 2\cdot 2^{4n\epsilon}\} \dot{\leq} 2^{n(R - D(T_X \| P_X))}.
\end{equation}
Using Lemma~\ref{lem:iid}, we obtain that $\mathbb{P}\{\varphi_{\mathcal{C}_n}(T_X)>2\cdot 2^{4n\epsilon}\}$ is doubly exponentially close to 0. Thus, we obtain
\begin{align}\label{eq:key-eta-1}
&\mathbb{E}_{\mathcal{C}_n}\sum_{y^{n}\in\mathcal{T}_{T_{Y}}}\phi_{\mathcal{C}_n(T_X)}^{1+s}(y^n)\nonumber\\
\leq& \mathbb{P}\{1\leq\varphi_{\mathcal{C}_n}(T_X)\leq 2\cdot 2^{4n\epsilon}\}2^{1+s}\cdot 2^{(2+s)4n\epsilon+nH(Y|X)_{T_{XY}}}+ \mathbb{P}\{\varphi_{\mathcal{C}_n}(T_X)>2\cdot 2^{4n\epsilon}\} 2^{nR}\nonumber\\
\dot{\leq}&2^{(2+s)4n\epsilon+nH(Y|X)_{T_{XY}}+n(R - D(T_X \| P_X))}
\end{align}
Then $\eta_1$ is upper bounded by
\begin{align}
\eta_{1}\dot{\leq}&\max_{T_{XY}:T_{X}\in\mathcal{T}_{1}}2^{-sn\sum T_{Y}\!\log\! P_{Y}+(1+s)n(\sum T_{XY}\!\log\! P_{Y|X}-R)}\cdot 2^{(2+s)4n\epsilon+nH(Y|X)_{T_{XY}}+n(R - D(T_X \| P_X))} \nonumber\\
=&\max_{T_{XY}:T_{X}\in\mathcal{T}_{1}}2^{snD(T_{Y}\|P_{Y})-(1+s)nD(T_{XY}\|P_{XY})+8n\epsilon}\cdot 2^{sn(D(T_{X|Y}\|P_{X}|T_Y)-R+4\epsilon)}, 
\end{align}

Next, we estimate $\eta_{2}$. By Lemma \ref{lem:packing-coveing-IID},
there is a realization $c'$ of $\mathcal{C}_{n}$ satisfying $\mathcal{B}_{1}(\epsilon|P_{X})\cap\mathcal{B}_{2}(\epsilon|P_{X})\cap\mathcal{B}_{3}(\epsilon|P_{X})$. Let $\gamma$ be defined in Eq.~\eqref{eq:gamma}. So, $\gamma$ is upper bounded by the maximum
of $\gamma_{1}'$ and $\gamma_{2}'$ defined below.
\begin{align}
\gamma_{1}'  :=&\max_{T_{X}\in\mathcal{T}_2,T_{Y|X}\in\mathcal{T}_3}\sum_{y^{n}\in\mathcal{T}_{T_{Y}}}2^{-sn\sum T_{Y}\!\log\! P_{Y}+(1+s)n(\sum T_{XY}\!\log\! P_{Y|X}-R)}\cdot2^{n(1+s)(R_{T_{X}}-I(X:Y)_{T_{XY}}+3\epsilon)} \nonumber\\
\dot{\leq}&\max_{T_{X}\in\mathcal{T}_2,T_{Y|X}\in\mathcal{T}_3} 2^{nH(Y)_{T_{Y}}-sn\sum T_{Y}\!\log\! P_{Y}+(1+s)n(\sum T_{XY}\!\log\! P_{Y|X}-D(T_{X}\|P_{X})-I(X:Y)_{T_{XY}}+3\epsilon)}\label{eq:gamma-1}\\
=&\max_{T_{X}\in\mathcal{T}_2,T_{Y|X}\in\mathcal{T}_3} 2^{snD(T_{Y}\|P_{Y})-(1+s)n(D(T_{XY}\|P_{XY})-3\epsilon)}\label{eq:sc-gamma1}
\end{align}
and 
\begin{align}
\gamma_{2}' &:=\max_{T_{X}\in\mathcal{T}_2,T_{Y|X}\in\mathcal{T}_4} \sum_{y^{n}\in\mathcal{T}_{T_{Y}}}2^{-sn\sum T_{Y}\!\log\! P_{Y}+(1+s)n(\sum T_{XY}\!\log\! P_{Y|X}-R)}2^{(1+s)5n\epsilon}\cdot\boldsymbol{1}\Big\{y^{n}\in\bigcup_{x^{n}\in c'(T_{X})}\mathcal{T}_{T_{Y|X}}(x^{n})\Big\} \nonumber\\
& \leq\max_{T_{X}\in\mathcal{T}_2,T_{Y|X}\in\mathcal{T}_4} 2^{-sn\sum T_{Y}\!\log\! P_{Y}+(1+s)n(\sum T_{XY}\!\log\! P_{Y|X}+5\epsilon)-(1+s)nR+n(R_{T_{X}}+H(Y|X)_{T_{XY}})}\label{eq:-2-1}\\
&\doteq \max_{T_{X}\in\mathcal{T}_2,T_{Y|X}\in\mathcal{T}_4} 2^{snD(T_{Y}\|P_{Y})-(1+s)n(D(T_{XY}\|P_{XY})-5\epsilon)}\cdot2^{sn(D(T_{X|Y}\|P_{X}|T_Y)-R)}, \label{eq:sc-gamma2-1}
\end{align}
where Eq.~\eqref{eq:gamma-1} is due to Eqs.~\eqref{eq:type-size} and \eqref{eq:sc-RTX}, Eq.~\eqref{eq:-2-1} follows from the inequality
\begin{equation}
\Big|\bigcup_{x^{n}\in c'(T_{X})}\mathcal{T}_{T_{Y|X}}(x^{n})\Big|\leq 2^{n(R_{T_{X}}+H(Y|X)_{T_{XY}})},
\end{equation}
and Eq.~\eqref{eq:sc-gamma2-1} comes from Eq.~\eqref{eq:sc-RTX}. Combining Eqs.~\eqref{eq:sc-gamma1} and \eqref{eq:sc-gamma2-1} gives
\begin{align}
\gamma\dot{\leq}\max_{T_{XY}:T_{X}\in\mathcal{T}_{2}}2^{snD(T_{Y}\|P_{Y})-(1+s)n(D(T_{XY}\|P_{XY})-5\epsilon)}\max\{1,2^{sn(D(T_{X|Y}\|P_{X}|T_Y)-R+4\epsilon)}\}.
\end{align} 
Lemma \ref{lem:packing-coveing-IID} shows that with probability doubly exponentially close to $1$, $\mathcal{C}_{n}$
takes such realizations $c'$. Thus, it holds that
\begin{equation}
\eta_{2}\dot{\leq}\max_{T_{XY}:T_{X}\in\mathcal{T}_{2}}2^{snD(T_{Y}\|P_{Y})-(1+s)n(D(T_{XY}\|P_{XY})-5\epsilon)}\max\{1,2^{sn(D(T_{X|Y}\|P_{X}|T_Y)-R+4\epsilon)}\}.
\end{equation}
\end{IEEEproof}

\section{Miscellaneous Lemmas}
\label{app:Miscellaneous-Lemmas} \setcounter{equation}{0} 
\global\long\def\theequation{B.\arabic{equation}}%
This appendix contains several technical lemmas that are used in the proofs. 
\begin{lemma}\label{lem:RenyiD-properties} 
Let $P$ and $Q\in\mathcal{P}(\mathcal{X})$. Then the R{\'e}nyi divergence
satisfies the following properties:
\begin{enumerate}
\item Monotonicity w.r.t. the order~\cite{VanHarremos2014renyi}: If $0\leq\alpha\leq\beta$, then $D_\alpha(P\|Q)\leq D_\beta(P\|Q)$.
\item Variational expression~\cite{CsiszarMatus2003information}: The R{\'e}nyi divergence can be written as
\begin{equation}
D_{\alpha}(P\|Q)=
\begin{cases}
\min\limits_{S\in\mathcal{P}(\mathcal{X})}\left\{\frac{\alpha}{1-\alpha}D(S\|P)+D(S\|Q)\right\}, & \alpha\in(0,1)\\
\max\limits_{S\in\mathcal{P}(\mathcal{X})}\left\{\frac{\alpha}{1-\alpha}D(S\|P)+D(S\|Q)\right\}, & \alpha\in(1,\infty).
\end{cases}
\end{equation}
\item Data processing inequality~\cite{AliSilvey1966general}: Let $\mathcal{W}:\mathcal{X}\rightarrow\mathcal{Y}$
be a channel. For any $\alpha\in[0,\infty]$, we have
\begin{equation}
D_{\alpha}(\mathcal{W}(P)\|\mathcal{W}(Q))\leq D_{\alpha}(P\|Q),
\end{equation}
where $\mathcal{W}(P):=\sum_x\mathcal{W}(\cdot|x)P(x)$ denotes the output distribution of $\mathcal{W}$ given input distribution $P$.
\end{enumerate}
\end{lemma}

\begin{lemma}[Minkowski's Inequality \cite{Rudin1976principles}]
Let $V_X$ and $W_X$ be vectors on a finite set $\mathcal{X}$. Then
\begin{align}
\|V_X + W_X\|_p &\le \|V_X\|_p + \|W_X\|_p, \quad p \ge 1 \\
\|V_X + W_X\|_p &\ge \|V_X\|_p + \|W_X\|_p, \quad p \in (0,1).
\end{align}
\end{lemma}

\begin{lemma}[\cite{CoverThomas1991elements}] \label{lem:-2}
Assume that $\left\{ a_{i}\right\} $
are non-negative real numbers. Then for $p\geq1$, we have
\begin{align}
\sum_{i}a_{i}^{p}\leq&\Big(\sum_{i}a_{i}\Big)^{p}, \quad p \ge 1 \\
\sum_{i}a_{i}^{p}\geq&\Big(\sum_{i}a_{i}\Big)^{p}. \quad p \in (0,1).
\end{align}
\end{lemma}

\begin{lemma}\label{lem:concave}
Let $0 \leq x,y \leq 1$ and $x+y \le 1$. For any $a,b\in(0,\infty)$, define 
$f(a,b)=a^xb^y$. Then $f$ is jointly concave on $(0,\infty)^2$.
\end{lemma}
\begin{IEEEproof}
To prove the joint concavity, it suffices to show that for any $ (a_1, b_1), (a_2, b_2) \in (0,\infty)^2$ and any $ \theta \in [0, 1] $, the following inequality holds:
\begin{equation}\label{eq:concave-desired}
(\theta a_1 + (1-\theta) a_2)^x (\theta b_1 + (1-\theta) b_2)^{y} \geq \theta a_1^x b_1^{y} + (1-\theta) a_2^x b_2^{y}.
\end{equation}
Let $ A = \theta a_1 + (1-\theta) a_2 $ and $ B = \theta b_1 + (1-\theta) b_2 $. Define
\begin{equation}
u_1 = \frac{a_1}{A}, \quad u_2 = \frac{a_2}{A}, \quad v_1 = \frac{b_1}{B}, \quad v_2 = \frac{b_2}{B}.
\end{equation}
From these definitions, we obtain that
\begin{align}
\theta u_1 + (1-\theta) u_2 &= \theta \frac{a_1}{A} + (1-\theta) \frac{a_2}{A} = 1, \label{eq:const1} \\
\theta v_1 + (1-\theta) v_2 &= \theta \frac{b_1}{B} + (1-\theta) \frac{b_2}{B} = 1. \label{eq:const2}
\end{align}
The right-hand side of Eq.~\eqref{eq:concave-desired} can be rewritten as:
\begin{align}
&\theta a_1^x b_1^{y} + (1-\theta) a_2^x b_2^{y} \nonumber\\
=& \theta (u_1 A)^x (v_1 B)^{y} + (1-\theta) (u_2 A)^x (v_2 B)^{y} \nonumber\\
=& A^x B^{y}[ \theta u_1^x v_1^{y} + (1-\theta) u_2^x v_2^{y}].
\end{align}
Thus, Eq.~\eqref{eq:concave-desired} is equivalent to
\begin{equation}
A^x B^{y} \geq A^x B^{y} \left[ \theta u_1^x v_1^{y} + (1-\theta) u_2^x v_2^{y} \right],
\end{equation}
which simplifies to
\begin{equation}\label{eq:target}
1 \geq \theta u_1^x v_1^{y} + (1-\theta) u_2^x v_2^{y}. 
\end{equation}
For each $ i = 1, 2 $, applying the weighted AM-GM inequality with weights $ x $ and $ y $, we have
\begin{equation}
u_i^x v_i^{y}\cdot1^{1-x-y} \leq x u_i + y v_i+1-x-y.
\end{equation}
Therefore,
\begin{equation}\label{eq:concave-1}
\theta u_1^x v_1^{y} + (1-\theta) u_2^x v_2^{y} \leq \theta \left( x u_1 + y v_1 \right) + (1-\theta) \left( x u_2 + y v_2 \right)+1-x-y.
\end{equation}
Substituting Eqs.~\eqref{eq:const1} and \eqref{eq:const2} into Eq.~\eqref{eq:concave-1}, we obtain
\begin{align*}
\theta( x u_1 + yv_1 ) + (1-\theta) ( x u_2 + y v_2)+1-x-y=1.
\end{align*}
This confirms Eq.~\eqref{eq:target}. The joint concavity follows.
\end{IEEEproof}

\begin{lemma}
[\cite{MitzenmacherUpfal2005probability}] \label{lem:chernoff-bound}
If $X^{k}$ is a sequence of i.i.d. Bern$(p)$ random variables with
$0\leq p\leq1$, then for $0<\delta<1$, 
\begin{equation}
\mathbb{P}\Big[\Big|\sum_{i=1}^{k}X_{i}-kp\Big|\geq\delta kp\Big]\leq2^{-\frac{\delta^{2}kp}{3}}.
\end{equation}
Here, a Bern$(p)$ random variable takes value $1$ with probability $p$ and $0$ with probability $1-p$.
\end{lemma}

\begin{lemma}
[Sion's Minimax Theorem~\cite{Sion1958general}]\label{lem:minimax-Thm} 
Let $\mathcal{A}$
be a compact convex set in a topological vector space $\mathcal{V}$
and $\mathcal{B}$ be a convex subset of a vector space $\mathcal{U}$.
Let $f:\mathcal{A}\times\mathcal{B}\rightarrow\mathbb{R}$ be such
that 
\begin{enumerate}
\item $f(a,\cdot)$ is quasi-concave and upper semi-continuous on $\mathcal{B}$
for each $a\in\mathcal{A}$, and 
\item $f(\cdot,b)$ is quasi-convex and lower semi-continuous on $\mathcal{A}$
for each $b\in\mathcal{B}$. 
\end{enumerate}
Then, we have 
\begin{equation}
\inf_{a\in\mathcal{A}}\sup_{b\in\mathcal{B}}f(a,b)=\sup_{b\in\mathcal{B}}\inf_{a\in\mathcal{A}}f(a,b),\label{eq:min-max}
\end{equation}
and the infima in Eq.~\eqref{eq:min-max} can be replaced by minima. 
\end{lemma}
%\bibliographystyle{IEEEtran}
%\bibliography{IEEEabrv,references}

\end{document}